\pdfoutput=1

\documentclass[a4paper,fleqn,usenatbib]{mnras}

\usepackage{txfonts}

\usepackage{amsmath}

\usepackage[T1]{fontenc}
\usepackage{ae,aecompl}

\usepackage{graphicx}	
\usepackage{amsmath}	
\usepackage{amssymb}	

\usepackage{color}
\usepackage{xcolor}
\usepackage{multirow}
\usepackage{lineno}
\usepackage[normalem]{ulem}

\definecolor{gris25}{gray}{0.25}
\definecolor{darkgreen}{rgb}{0,0.7,0.3}
\definecolor{darkblue}{rgb}{0.2,0,0.8}
\definecolor{pink}{rgb}{1.0, 0.05, 0.34}
\definecolor{lightblue}{rgb}{0.0, 0.5, 1.0}

\title[A new asteroseismic diagnostic for internal rotation in $\gamma$ Doradus stars]{A new asteroseismic diagnostic for internal rotation in $\gamma$ Doradus stars}

\author[R.-M. Ouazzani et al.]{
Rhita-Maria Ouazzani$^{1}$\thanks{E-mail: rhita-maria.ouazzani@phys.au.dk}, 
S\@.J\@.A\@.J\@. Salmon$^{2,3}$,
V\@. Antoci$^{1}$,
T\@.R\@. Bedding$^{4,1}$, \and
S\@.J\@. Murphy$^{4,1}$,
and I\@.W\@. Roxburgh$^{5}$
\\
~\\
$^{1}$Stellar Astrophysics Centre, Department of Physics and Astronomy, Aarhus University, Ny Munkegade 120, DK-8000 Aarhus C, Denmark,\\
$^{2}$Laboratoire AIM, CEA/DSM-CNRS-Universit\'e Paris 7, Irfu/Service d'Astrophysique, CEA-Saclay, 91191 Gif-sur-Yvette, France,\\
$^{3}$Institut d'Astrophysique et de G\'eophysique de l'Universit\'e de Li\`ege, All\'ee du 6 Ao\^{u}t 19, 4000 Li\`ege, Belgium,\\
$^{4}$Sydney Institute for Astronomy (SIfA), School of Physics, University of Sydney, Australia,\\
$^{5}$Astronomy Unit, Queen Mary University of London, Mile End Road, London, E1 4NS, UK.
}

\date{Accepted XXX. Received YYY; in original form ZZZ}

\pubyear{2016}

\begin{document}
\label{firstpage}
\pagerange{\pageref{firstpage}--\pageref{lastpage}}
\maketitle

\begin{abstract}
  With four years of nearly-continuous photometry from Kepler, we are finally in a good position to apply asteroseismology to $\gamma$ Doradus stars. In particular several analyses have demonstrated the possibility to detect non-uniform period spacings, which have been predicted to be directly related to rotation.
  In the present work, we define a new seismic diagnostic for rotation in $\gamma$ Doradus stars that are too rapidly rotating to present rotational splittings. Based on the non uniformity of their period spacings, we define the observable $\Sigma$ as the slope of the period spacing when plotted as a function of period. We provide a one-to-one relation between this observable $\Sigma$ and the internal rotation, which applies widely in the instability strip of $\gamma$ Doradus stars.
We apply the diagnostic to a handful of stars observed by {\it Kepler}.   
Thanks to g-modes in $\gamma$ Doradus stars, we are now able to determine the internal rotation of stars on the lower main sequence, which is still not possible for Sun-like stars.
\end{abstract}


\color{black}{}
\section{Introduction}

With four years of nearly-continuous photometry from Kepler, we are finally in a good position to apply asteroseismology to $\gamma$ Doradus ($\gamma$ Dor) stars. These stars are late A- to early F-type stars, with masses ranging from about 1.3 to 2 M$_{\odot}$. They burn hydrogen in their convective cores, surrounded by a radiative region, where a shallow convective layer appears. Such shallow convective envelopes trigger gravity oscillation modes (g-modes) driven by the convective blocking mechanism \citep[see][and references therein]{Guzik2000,Dupret2005}. These modes probe the innermost regions and in particular the interface between convective and radiative layers. Therefore, $\gamma$ Dor stars are ideal for studying convective core physics. It is expected that phenomen a such as transfer of angular momentum and mixing of chemical elements occur at the edge of convective cores, in the surrounding radiative regions \citep[see for instance][]{Kippenhahn1994}. Whether it is caused by convective overshooting, shear induced turbulence or gravity waves is still matter of debate.

Another interesting motivation for $\gamma$ Dor stars seismology lies in the fact that their mass range overlaps with the masses of the red giant stars  observed by the NASA {\it Kepler} space photometry mission. One of the most important developments in stellar physics that {\it Kepler} facilitated was the inference of the core rotation periods of red giant stars \citep{Beck2012,Deheuvels2012}. After 25 months of observation, from the seismic analysis of 1200 stars \cite{Mosser2012b} measured the core rotation periods of 300 red giants, among which over one third have masses between 1.2 and 2 M$_{\odot}$. These were shown to be in disagreement with stellar evolution modelling \citep{Eggenberger2012,Marques2013,Ceillier2013}. Since then, a number of non-standard angular momentum transport processes have been implemented in stellar evolution models: transport by meridional circulation and shear induced turbulence \citep{Eggenberger2012,Marques2013}, transport by the Taylor-Spruit instability \citep{Heger2005,Cantiello2014}, by internal gravity waves \citep{Fuller2014} and transport by mixed modes \citep{Belkacem2015b}. However the slow core rotation of red giant stars at the base of the red giant branch still remains unexplained by theory. For that reason, we lack knowledge of the inner rotation profile of low-to-intermediate mass stars on the main sequence (MS). The work presented here aims to address this. We establish a new method of measuring the core rotation periods of the intermediate-mass stars on the MS, specifically for $\gamma$ Doradus pulsators.

The g-modes in $\gamma$ Doradus stars typically have periods close to one day, which makes ground-based study exceedingly difficult. At first, the space missions such as MOST, CoRoT, and {\it Kepler} allowed to address the stability and hybridity of $\gamma$ Doradus stars \citep[see for instance][]{Grigahcene2010,Hareter2012,Zwintz2013,Uytterhoeven2011,Tkachenko2013}. After four years of nearly-continuous data from {\it Kepler}, it is now possible to perform detailed asteroseismic studies of $\gamma$ Dor stars based on their g-modes \citep[see][]{Bedding2015,VanReeth2015}. Among the {\it Kepler} stars with KIC temperatures between 6000 K and 10000 K, and $\log g$ values between 3.0 and 4.5, we have identified around 6000 A- early F-type stars. Approximately one thousand of these show excess power in the $\gamma$ Dor g-modes frequency range. 

 Unlike for low-mass stars, in $\gamma$ Dor stars the convective envelopes are too shallow to generate a magnetic field able to act as a magnetic torque and slow down their surfaces \citep{Schatzman1962}. For that reason, these stars typically have projected rotation velocities of around 100 km s$^{-1}$ but that can reach up to 250 km s$^{-1}$ \citep[see for instance][]{Abt1995,Royer2009}.

When a star rotates, the degeneracy of the frequencies is lifted and a rotational splitting appears between modes of same radial order ($n$) and angular degree ($\ell$) but different azimuthal order ($m$). If the star rotates slowly, the multiplet structure holds, and the splitting directly gives the average rotation rate in the cavity probed by the modes. This is possible thanks to a simple model of the effect of rotation as a first-order perturbation of oscillations, an approach derived by \cite{Ledoux1951}. Examples of measured rotation rates using that method can be found for instance in \cite{Kurtz2014}, \cite{Saio2015}, \cite{Schmid2015}, \cite{Keen2015}, and \cite{Murphy2016}. However most of {\it Kepler}'s $\gamma$ Dors do not show clear multiplet structures. Also, if the rotation period is larger or of the same order of the pulsation period, the rotational splitting becomes difficult to determine and no longer provides an accurate measure of the internal rotation rate. 

For moderately rotating stars, the traditional approximation of rotation \citep[hereafter TAR,][]{Eckart1960}, allows to account for rotation in the pulsation treatment. By making simplifying hypotheses, the TAR conserves the separability of the oscillation equations. In particular, it assumes the star is rotating uniformly. In that framework, \cite{Bouabid2013} showed the main features due to the effect of rotation on the seismic spectra of $\gamma$ Dor stars. With the full four years of data from {\it Kepler}, \cite{Bedding2015} and \cite{VanReeth2015} were able to identify such features in observations, opening the way to the seismic determination of rotation in $\gamma$ Doradus stars. The first attempt \citep{VanReeth2016} consisted of fitting the observed patterns to a grid of models using $\chi^2$-minimization. These pulsation models were calculated using the TAR. 

We opt for the use of seismic diagnostics which are model independent, and are therefore not affected by our lack of knowledge of the stellar structure. Our choice is to lift the simplifying hypotheses that are implied in the TAR, in order to obtain pulsation spectra of g-modes in rapidly rotating $\gamma$ Dor stars, with the quality required for comparison with observations. To do so, we make use of the non-perturbative pulsation modelling method developed by \cite{Ouazzani2012b,Ouazzani2015}. As shown in Sect. \ref{Ssec_2D}, the formalism used allows one to explore any kind of differential rotation profile, including rotational velocities close to the break-up velocity.

In Sect.\ref{Sec_Modelling} we give a brief description of the stellar models that have been explored in this study. In Sect. \ref{Sec_Osc} we present the non-perturbative approach, as well as of the traditional approximation of rotation, and we provide an in-depth comparison of the two methods to estimate the validity of the traditional approximation. In Sect. \ref{Sec_GlobalEffects}, we describe the global effects of rotation on the g-mode pulsation spectra and then, more specifically, on the g-modes period spacings. That allows us to define a new seismic diagnostic tool: the slope of the period spacing as a function of the period. We explore the impact of different parameters (i.e. stellar parameters, metallicity, centrifugal distortion, type of mixing) on that new diagnosis tool in Sect. \ref{Sec_SlopeEffects}. Finally, we discuss and summarize the results and give an exemple of application in Sect. \ref{Sec_Ccl}.

\section{Stellar Models}
\label{Sec_Modelling}
\begin{table*}
	\centering
	\caption{Parameters of the models explored in detail in this study. They have been name as follows:  the number 1, 2 3 corresponds to the mass (resp. 1.4, 1.6 and 1.86 M$_{\odot}$), then the letter z stands for ZAMS, m for mid main sequence, and t for TAMS, and no super script stands for models with turbulent diffusion (with Co = 700 cm$^2$ s$^{-1}$) and no overshooting, prime (') stands for no diffusion and no overshooting, and double prime ('') stands for no diffusion and overshooting (with $\alpha_{ov} = 0.2$). }
	\label{tab:models_table}
	\begin{tabular}{lccccccccr} 
		\hline
		Model name & 1z & 2m & 2z & 2t & 3t & 3m & 3m' & 3m'' \\
		\hline
		\, ~ \, $\rm M/M_{\odot}$ \, ~ \, & \, ~ \, 1.40 \, ~ \, & \, ~ \, 1.60 \, ~ \, & \, ~ \, 1.60 \, ~ \, & \, ~ \, 1.60 \, ~ \, & \, ~ \, 1.86 \, ~ \, & \, ~ \, 1.86 \, ~ \, & \, ~ \, 1.86 \, ~ \, & 1.86 \\
		\, ~ \, T$_{eff}$ \, ~ \, & 6880 & 7190 & 7785 & 6160 & 6760 & 7960 & 8025 & 7911 \\
		\, ~ \, $\log L/L_{\odot}$ \, ~ \, & 0.627 & 1.036 & 0.852 & 1.181 & 1.355 & 1.269 & 1.202 & 1.256 \\
		\, ~ \, $\log$ g \, ~ \, & 4.30 & 3.98 & 4.32 & 3.57 & 3.63 & 3.99 & 4.08 & 4.00 \\
		\, ~ \, $\rm R/R_{\odot}$ \, ~ \, & 1.38 & 2.13 & 1.46 & 3.43 & 3.48 & 2.27 & 2.07 & 2.26 \\
		\, ~ \, Age (Myr) \, ~ \, & 185 & 1830 & 174 & 2694 & 1480 & 1048 & 782 & 991 \\
		\, ~ \, X$_{c}$ \, ~ \, & 0.68 & 0.35 & 0.68 & 0.03 & 0.06 & 0.34 & 0.32 & 0.34 \\
                \, ~ \, D$_t$ (cm$^2$ s$^{-1}$) \, ~ \, & 700 & 700 & 700 & 700 & 700 & 700 & -- & -- \\
                \, ~ \, $\alpha_{ov}$ \, ~ \, & -- & -- & -- & -- & -- & -- & -- & 0.2 \\
		\hline
	\end{tabular}
\end{table*} 

Stellar models were computed with the stellar evolution code CLES \citep{Scuflaire2008b} for masses between 1.2 and 1.9 M$_{\odot}$, and with initial helium mass fraction $Y = 0.27$. We adopted the AGSS09 metal mixture \citep{Asplund2009} and corresponding opacity tables obtained with  OPAL opacities  \citep{Iglesias1996},  completed  at  low  temperature ($log T <4.1$) with \cite{Ferguson2005} opacity tables. We  used  the  OPAL2001  equation  of  state  \citep{Rogers2002}  and  the  nuclear  reaction  rates  from  NACRE  compilation \citep{Angulo1999}, except  for  the $^{14}N(p,\gamma) ^{15}O$  nuclear  reaction, for which we adopted the cross-section from \cite{Formicola2004}. Surface boundary conditions at $ T = T_{\rm eff}$ were provided by ATLAS model atmospheres \citep{Kurucz1998}. Convection was treated using the mixing-length theory (MLT) formalism \citep{Bohm-Vitense1958}  with a parameter $\alpha_{\textrm{MLT}} = 1.70$.

We considered models with and without instantaneous convective core overshooting (expressed in local pressure scaleheight, $H_{\textrm{p}}$), as well as with and without turbulent diffusion. Since the CLES code does not include effects of rotation on transport of angular momentum or chemical species, we instead introduced mixing by turbulent diffusion, following the approach of \citet{Miglio2008}. This reproduces an effect of rotationaly-induced mixing that is quite similar to overshooting, but in addition tends to smooth chemical composition gradients inside the star. In models including this type of mixing, the coefficient of turbulent diffusion was set to $D_{\textrm{t}}=700 cm^2/s$ and kept constant to this value during evolution and in every layer of the models. This value was selected from a previous calibration to Geneva models with similar masses \citep{Miglio2008}.

A grid of stellar models was computed accordingly to these prescriptions, whose parameters are given in Table \ref{tab:models_grid}. We did not calculate pulsational stability, but we made sure that the grid maps the entire $\gamma$ Doradus instability domain \citep{Bouabid2013}. Table \ref{tab:models_table} gives more details of the parameters of specific models used below for illustration purposes.

As explained by \cite{Miglio2008}, if no specific mixing is added to models with growing convective cores (and therefore a chemical discontinuity) on the main sequence, a contradictory situation appears in regards to the convective criteria, leading to a semi-convective region. More generally, the stratification at the boundary of the convective core depends on the numerical treatment adopted. In our CLES version, we do not apply a specific treatment to these transition layers. Instead, a double mesh point is introduced at the boundary of the convective core in the models used for the computations of oscillations. This required modification of part of our pulsation codes, i.e. the implementation of matching conditions at interfaces between the convective core and the surrounding radiative region. This treatment is specially important for pulsational studies, to prevent numerical artefacts creating additional cavities trapping the g-modes.  


    \begin{table}
  \centering
  \caption{ Parameters of the grid of models used in this work.
  }
  \label{tab:models_grid}
  \begin{tabular}{lcccr} 
		\hline
		quantity \, ~ \, & \, ~ \, values \, ~ \, & \, ~ \, step  \\
                \hline
		$\rm M/M_{\odot}$ \, ~ \, & \, ~ \, 1.20 - 1.90 \, ~ \, & \, ~ \, 0.02 \\
		$Z$ \, ~ \, &  \, ~ \, 0.010- 0.018 \, ~ \, & \, ~ \, 0.004 \\
                $\alpha_{\rm MLT}$ \, ~ \, & \, ~ \, 1.70 \, ~ \, & \, ~ \, -- \\
                $\nu_{\rm rot} \rm \,(\mu \rm Hz)$ \, ~ \, & \, ~ \, 0.0 \, - \, 23.0 \, ~ \, & \, ~ \, 2.0 \\
                $\rm D_t \,(cm^2 s^{-1})$ \, ~ \, & \, ~ \, 0 \, / \, 700 \, ~ \, &\, ~ \,  -- \\
                $\alpha_{ov}$ \, ~ \, & \, ~ \, 0 \, / \, 0.2 \, ~ \, & \, ~ \, -- \\
                \hline
  \end{tabular}
  \end{table}

\section{Oscillation computation methods}
\label{Sec_Osc}

Rotation induces centrifugal and Coriolis forces which distort the star, i.e. the equilibrium structure is no longer spherical, and modify the spectrum of eigenfrequencies and eigenfunctions of the oscillation modes. In this article, we explore gravity and gravito-inertial pulsation modes of $\gamma$ Doradus stars models. In rotating stars, these modes are often computed assuming simplifying hypotheses. This is the case with the traditional approximation of rotation (TAR, developed for geophysics by \citealt{Eckart1960}, and applied to stellar pulsations by \citealt{Berthomieu1978}). It neglects the centrifugal distortion and part of the contributions  of the Coriolis force. Another approach is to fully account for the Coriolis force while neglecting the centrifugal distortion \citep{Dintrans2000}, or while acounting for the distortion by means of the second-order Legendre polynomial \citep[i.e., using a Chandrasekhar and Milne expansion, see][]{Lee1995}\color{black}{}. The most complete approach accounts for both the Coriolis force and the full effect of the centrifugal force through a two-dimensional (2D) non-perturbative method as developed for polytropes by \cite{Reese2006}, for non polytropic-homogeneous models by \cite{Reese2009a}, and for distorted models of $\delta$ Scuti stars by \cite{Ouazzani2015}. These three methods --i.e. TAR, complete one-dimensional (1D) and complete 2D-- have been compared by \cite{Ballot2012} using polytropic models. According to them, the 1D non-perturbative approach -- which presents the advantage of requiring less numerical ressources --, and the TAR give satisfactory results compared to the full 2D approach. Throughout this study we therefore work with one-dimensional spherical stellar models. A further justification for this is given in Sect. \ref{Ssec_2D}. Moreover, in this article we have performed calculations with both the TAR and the non-perturbative approach. This section describes the two methods in more detail, and comparing them for spherical models of $\gamma$ Doradus stars.

\subsection{Non-perturbative Modelling}
\label{Ss_ACOR}

The oscillation modes were computed as the adiabatic response of the structure to small perturbations,  i.e. of the density, pressure, gravitational potential, and velocity field. The  ACOR code was developed for this purpose and is presented in \cite{Ouazzani2012b}.
This oscillation code solves the hydrodynamics equations perturbed by Eulerian fluctuations, performing direct integration of the 2D problem. The numerical method is based on a spectral multidomain method which expands the angular dependence of eigenfunctions into spherical harmonics series, and whose radial treatment is particularly well adapted to the behaviour of equilibrium quantities in evolved models (at the interface of convective and radiative regions, and at the stellar surface). The radial differentiation scheme is made by means of a sophisticated finite difference method, which is accurate up to the fifth order in terms of the radial resolution \citep[developed by][for the LOSC adiabatic code]{Scuflaire2008a}.
This code has been validated by comparison with the results of \cite{Reese2006} for polytropic models. The agreement between the two codes is excellent. 

In this study, we have relied on the non-adiabatic stability calculations provided in \cite{Bouabid2013} in order to determine the range of radial orders to investigate. All the pulsations spectra presented here, whatever the method used, are calculated for radial orders between $n=-50$ to $-20$. Note that, by convention, g-modes radial orders are negative. With the non-perturbative approach in particular, we have calculated pulsation spectra of dipolar modes for all the models whose parameters are given in Tab. \ref{tab:models_table}, and for quadrupolar modes for the models 1z, 2m and 3t, covering a range of uniform rotation frequencies, from zero to 23 $\mu$Hz.

Here, the pulsation modes have been computed using up to 5 spherical harmonics, i.e. for a given $m$,\, $\ell$=1 to 9 with odd value for dipolar modes, and $\ell$=2 to 10 with even value for the quadrupolar ones except for $\ell$=2 zonal modes, where the $\ell$=0 component is included as well). We have made sure, by the mean of convergence tests, that the frequencies do not vary significantly when adding a sixth spherical harmonic in the series. The resulting modes have multiple $\ell$ characters, they are then assigned an effective angular degree, taking the dominant contribution in the series. 

\subsection{The traditional approximation of rotation}
\label{Ss_TAR}

In rotating stars, the equation system for pulsations is not separable in terms of the radial and horizontal coordinates, unlike for the case without rotation. The TAR is an approximate treatment that conserves the separability of the system. The first hypothesis is made on the rotation profile by assuming solid-body rotation. The centrifugal distortion is neglected and hence spherical symmetry is assumed. Furthermore, considering the properties of low frequency high order g-modes, the TAR neglects the Coriolis force associated with radial motion and radial component of the Coriolis force associated with horizontal motion. Practically, it consists of neglecting the horizontal component of the angular velocity so that ${\bf\Omega} = \left[\Omega \cos \theta, 0, 0 \right]$ in the spherical polar coordinates. Finally, the Cowling approximation is made \citep{Cowling1941}, which neglects the perturbation of the gravitational potential. As a result, the motion equation for pulsations is separable in terms of the radius and latitude. It can be reduced to an equation for the radial component, which is similar to the one without rotation, and a Laplace tidal equation for the horizontal component, whose eigenfunctions are the Hough functions. For a detailed derivation of these equation, we refer to \cite{Unno1989} Sect. 34.3, or \cite{Lee1987a}. 

This approximation has been implemented in the LOSC adiabatic and MAD non-adiabatic oscillation codes, the details of which have been given by \cite{Bouabid2013} and \cite{Salmon2014}. 

We have computed a selection of spectra with this method, for the sake of comparison with the non-perturbative approach, for the models 1z, 2m, and 3m'', and for the whole rotation frequency range from 0 to 23 $\mu$Hz.

\subsection{Asymptotic formulation of the TAR}
\label{Ss_AsympTAR}

Under the TAR approximation, the simplification of the problem allows for an asymptotic formulation derived from the \cite{Tassoul1980} formula for g-modes periods, where $\ell (\ell+1)$ is replaced by $\lambda$. This eigenvalue depends on $\ell$, $m$ and the spin parameter $s = 2 \nu_{\rm rot} / \nu_{\rm co}$, $\nu_{\rm co}$ being the frequency of the mode in the corotating frame. For the sake of clarity we will use hereafter $\lambda_{\ell,m,s(n)}$, with $n$ the radial order of the mode. The asymptotic formula is then given by:
\begin{equation}
  \rm P_{\rm co}(n)= \frac{2\pi^2 (n+\frac{1}{2})}{\sqrt{\lambda_{\ell,m,s(n)}}\int_{r_0}^{r_1}\frac{N}{r} \rm dr},
\end{equation}
where r$_0$ and r$_1$ are the limits of the g-modes cavity, determined by $\omega^2 < \rm N^2, S_{\ell}^2$, i.e. for an angular frequency squared smaller than both the square of the Brunt-Vaissala ($N$) frequency, and the square of the Lamb frequency ($S_{\ell}$). The period spacing, given as the difference between the periods of two modes of same angular degree, same azimuthal order, and consecutive radial orders, after some calculations \citep[see Appendix A of][]{Bouabid2013}, can be expressed as follows: 
\begin{align}
  \label{TARAsympt}
  \langle \Delta \rm P_{\rm co}\rangle \simeq \frac{2\pi^2}{\sqrt{\lambda_{\ell,m,s(n+1)}}\int_{r_0}^{r_1}\frac{N}{r} \rm dr\left( 1 + \frac{1}{2} \frac{\rm d \ln \lambda}{\rm d \ln s} \right) } .
\end{align}
  Note that the period appears in both terms (hidden in the spin parameter $s$) of these equations in non-linear terms, the periods and period spacings are therefore obtained using an iterative scheme. Initial guesses of the periods were obtained by taking the adiabatic periods obtained in the case without rotation (low numerical cost) as input. The equation was then solved with the Van Wijngaarden-Dekker-Brent iterative scheme, which is a refined bisection method \citep{Press1996}.
  
The asymptotic formulation requires very little computational resources and time, which is an clear advantage compared to the aforementioned methods. For that reason, it is used in this study for a wide exploration of the HR diagram (see Sect. \ref{Ssect:stelpar}). However, we first investigate its domain of validity by comparing with the non-perturbative method and the TAR. 

\subsection{Comparison of methods}
\label{Ss_comparaison_ACOR-TAR}
 \begin{figure}
  	\includegraphics[width=1.\linewidth]{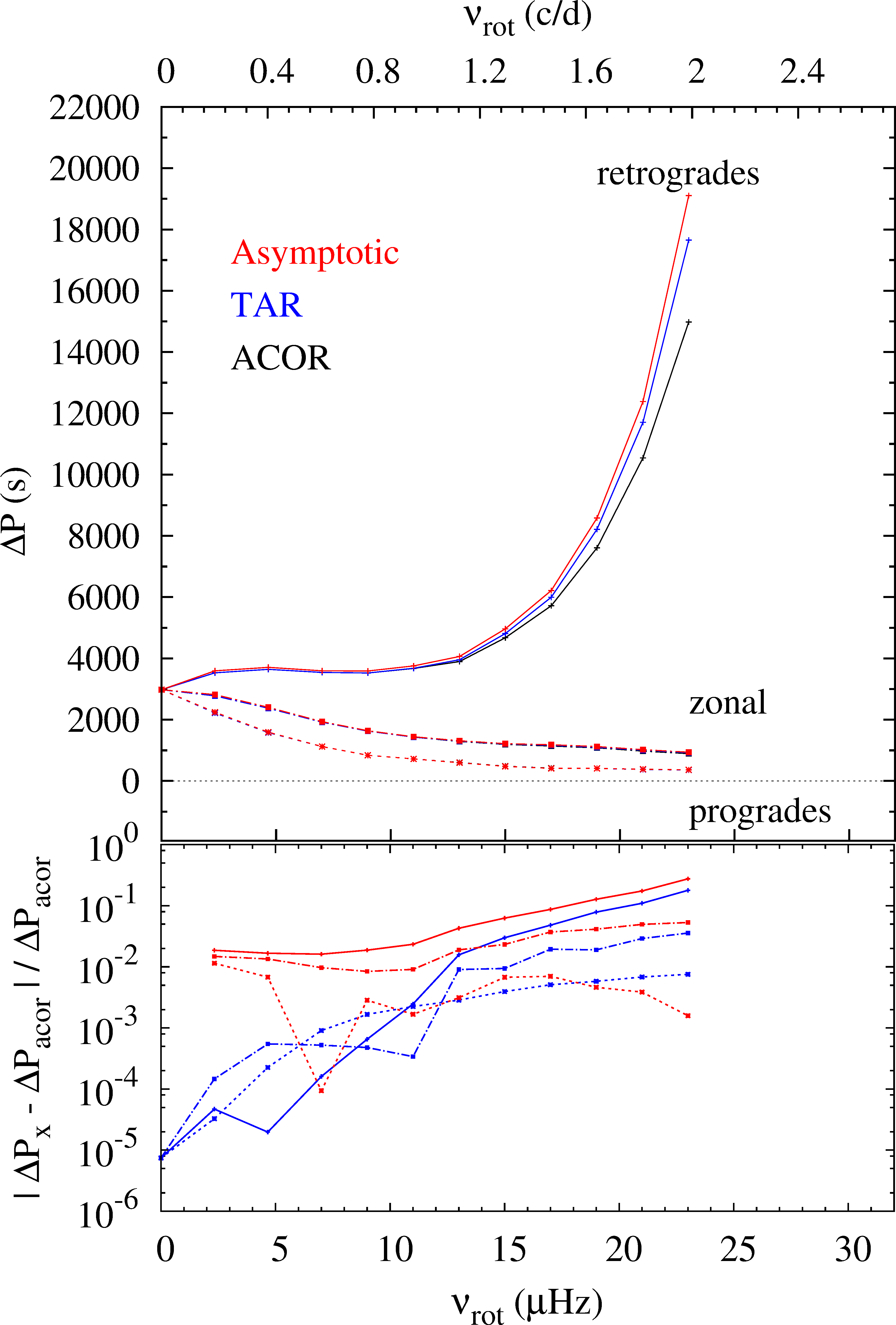}
        \caption{{\it Top}: Mean period spacing for modes with radial orders ranging from $n=-20$ to $n=-50$, computed with the non-perturbative method (ACOR, black), the traditionnal approximation (TAR, blue), and the Asymptotic formula derived from the traditional approximation (Asymptotic, red), for model 1z given in Tab. \ref{tab:models_table}. {\it Bottom}: relative discrepancies with respect to the non-perturbative calculations for the TAR (blue), and for the Asymptotic formulation (red). The solid lines correspond to the retrograde modes, the dot-dashed lines to the zonal modes, and the dotted lines to the prograde modes.}
    \label{fig:DeltaPm_S33}
\end{figure}

\begin{figure*}
\centering
	\includegraphics[scale=1]{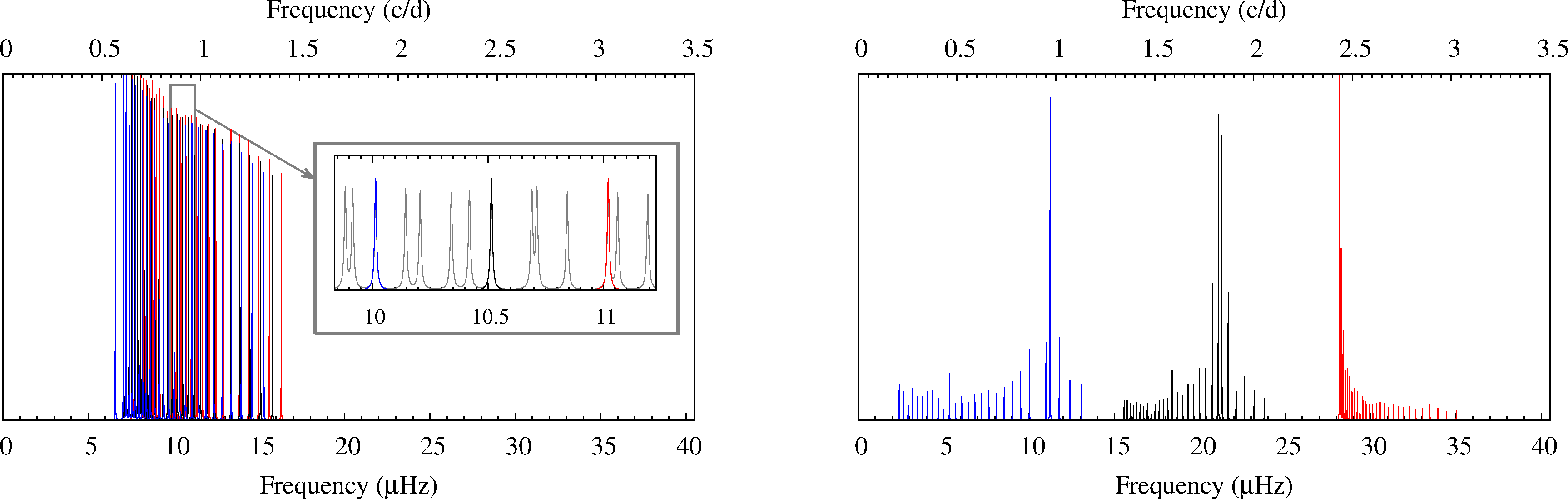}
    \caption{Frequency spectra for model 2m (see Tab. \ref{tab:models_table}), {\it left:} Rotating at a period of 11.6 days,  {\it right:} Rotating at a period of 0.5 days. On the vertical axis has been plotted the inertia to the power $-1/4$. In black are the quantities relative to zonal modes, in red to the prograde modes, and blue to the retrograde ones. In the zoom box on the left panel, the triplet {$\ell=1, n=-32$} is shown among the $\ell=1$ modes spectrum. The quantity plotted in the y-axis is the inverse of modes inertia, although this is not relevant here.}
    \label{fig:spect}
\end{figure*}

It is important to emphasize that the choice of the set of variables used in the pulsation calculations can lead to different numerical results \citep[see e.g.][]{Godart2009}. The non-perturbative method implemented in ACOR relies on the Eulerian formalism, whereas for the TAR, one can choose between the Eulerian or the Langrangian formulations. In the case where different choices are made for the two calculations, even without rotation, the discrepancy between the resulting pulsation frequencies is as high as 2$\%$ in relative value (approximately 0.3$\mu$Hz at 15.0$\mu$Hz, i.e. 0.03 c/d at 1.30 c/d). This is due to the fact that the Brunt-Va\"iss\"al\"a frequency ($N^2$) does not appear explicitely in the Lagrangian formulation of the motion equation. Due to numerical truncation errors, $N^2$ can wrongly take positive or negative values at the edge of a convective region. The fact that $N^2$ appears explicitely in the Eulerian formulation of the equations ensures that the correct value is assigned.
We confirm the statement in \cite{Godart2009} for $\gamma$ Doradus stars, which present one or several convective regions: the Eulerian formulation of the pulsation equations is the most appropriate to the finite difference scheme. Accordingly, the Eulerian variables are used hereafter in both ACOR and the TAR.

We calculated the discrepancy between the frequencies given by ACOR and by the TAR in the non-rotating limit, for Eulerian variables. For instance, for model 1z (parameters in Tab.\ref{tab:models_table}), the discrepancies are at most 3$\times 10^{-3} \%$ in relative value. It is of the same order of magnitude for models 2m and 3m''.

For comparison between the three methods, we investigated the value of the mean period spacing.  In the asymptotic method, it is given by Eq.\ref{TARAsympt}, using the simple change of reference frame, possible only in the case of uniform rotation:
\begin{align}
  \rm P_{\rm in} = \frac{P_{\rm co}}{ 1\,- \,m \frac{P_{\rm co}}{P_{\rm rot}}}\,.
  \label{eq:Pin}
\end{align}
For the TAR and the non-perturbative approach, we simply compute the mean period spacing as the average of the period spacings over the eigenmodes with radial orders between $-50$ and $-20$. Figure \ref{fig:DeltaPm_S33} gives the mean period spacing as a function of the uniform rotation frequency computed with the three methods in the top pannel. The largest discrepancies arise for the retrograde modes (solid lines), whereas the zonal and prograde modes agree well. In the bottom pannel, we give the discrepancy of the TAR and of its asymptotic formula, relative to ACOR. For retrograde modes, the error on the mean period spacing is of 15\% for the asymptotic values. This corresponds to a discrepancy of 2600s (0.03 days) with the TAR and 4300s (0.05 days) with the asymptotic formula for a value of the mean period spacing around 17 000s (0.2 days). However, it seems that, for any azimuthal order, the TAR is giving values that are in agreement with the complete calculations within 0.2\% for rotation frequencies lower than 11 $\mu$Hz (1 c/d), whereas they agree within 2\% for the asymptotic ones. For rotation frequencies higher than 11$\mu$Hz, the retrograde modes period spacings computed with the TAR differ from the non-perturbative ones by 7\% in the worst case. The asymptotic formulation is more problematic and can sometimes reach errors of around 15 \% on the mean period spacing. 

The worst case arises for retrograde modes. The impact of the number of spherical harmonics included in the pulsation modelling has been investigated, and can be excluded as sources for these discrepancies.
In fact, when calculated in the corotating frame, the discrepancies between the mean period spacings given by the three methods are of same order for the different $m$ components. The stronger discrepancies in the inertial frame are to be attributed to the denominator in Eq. \ref{eq:Pin}, which is small when pulsation periods in the corotating frame are close to the rotation period. Because retrograde (resp. prograde) modes are shifted towards shorter periods in the corotating frame, i.e. closer to the rotation period, the denominator in Eq. \ref{eq:Pin} is smaller for retrograde modes than for prograde modes (that are shifted towards longer periods). For example, for the highest rotation rate in Fig. \ref{fig:DeltaPm_S33} ($\nu_{\rm rot} = 23 \mu$ Hz, P$_{\rm rot} = 0.5$ days), the excited modes have periods between 0.3 and 0.5 days for retrograde modes, and 0.9 and 2.3 days for prograde modes. Therefore, the discrepancies are enhanced by the change of reference frame for retrograde modes. Hence the need to adopt the non-perturbative method when dealing with observations of retrograde modes.

\color{black}{}

\section{Effect of Rotation on g-modes}
\label{Sec_GlobalEffects}

\subsection{Global effect on the spectrum}
\label{Ss_spectrum}

    Here we explore the global effects of rotation on the frequency spectrum. We aim to give some insights for interpretation of observed pulsation spectra, giving the main indicators in terms of observables.

Even in the corotating frame, and even when neglecting the distortion due to the centrifugal force, the degeneracy of pulsation modes is lifted, and the rotational splitting of frequencies appears. 
As the rotation rate increases, three rotation regimes can be identified. The classification of these regimes depends on the magnitude of the rotational splitting $\delta \nu_{m}$ (i.e. the frequency difference between modes of the same $n, \ell$ and different m) compared to the frequency separation between the modes of same $\{\ell, m\}$ and consecutive radial orders $\Delta \nu_n$. Because g-modes are supposed to be equally spaced in period rather than in frequency, we can express that frequency separation using the period spacing ($\Delta \nu_n \simeq \Delta \rm P / P^2$). Therefore we consider the factor $\delta \nu_{m} / (\Delta \rm P / P^2)$ as the relevant indicator of the effect of rotation on g-modes spectrum:
\begin{itemize}
\item[-] For slow rotation, $\delta \nu_{m} / (\Delta \rm P / P^2) < 10^{-1}$, i.e. the rotational splitting is at least one order of magnitude smaller than the frequency separation. In this case, the split multiplets can be easily identified, and the rotation rate can be derived from the rotational splitting using a linear formulation \citep{Ledoux1951}. The period spacing is not significantly sensitive to rotation, and remains the same for the different components of the multiplets.
\item[-] For moderate rotation, $\delta \nu_{m} / (\Delta \rm P / P^2) \gtrsim 1$, the splitting can be larger than the frequency separation. Therefore, the multiplets are more difficult to identify in the spectrum, but they remain possible to highlight by using different visualization tools such as for instance {period \'echelle diagrams}. See \cite{Bedding2015} for the first use of such a tool for {\it Kepler}'s $\gamma$ Dor, or \cite{Keen2015} for a specific example. We give an illustration of a frequency spectrum in this regime in Fig. \ref{fig:spect} (left).
\item[-] For rapid rotation, $\delta \nu_{m} / (\Delta \rm P / P^2) \gg 1$, the structure of the pulsation spectrum changes radically. The prograde modes are shifted towards higher frequency, whereas the retrogrades are shifted towards lower frequencies, to such an extent that they appear in the frequency spectrum as {\it clusters} of modes, each with a given angular degree and azimutal order and varying radial order (see Fig. \ref{fig:spect}, right  and \citealt{VanReeth2015} for such features in {\it Kepler} observations). Note that this frequency grouping due to rapid rotation has been found also in a Be stars observed by MOST \citep{Cameron2008}.
\end{itemize}

\subsection{Effect of rotation on the period spacing}
\label{Ss_periodspacing}
\begin{figure}
  \centering
	\includegraphics[scale=0.9]{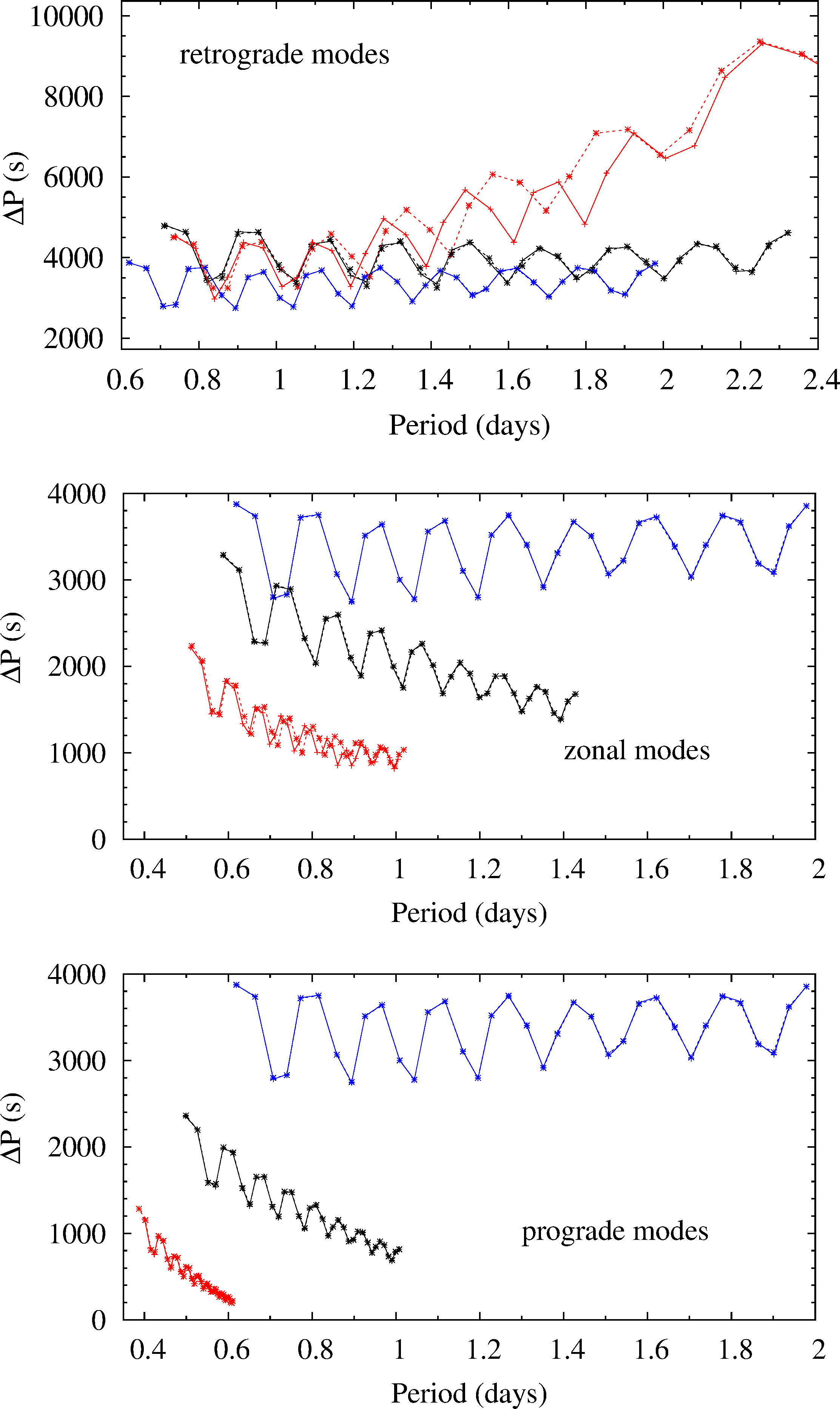}
        \caption{Period spacing as a function of the period for dipolar ($\ell=1$) retrograde ($m=+1$, top), zonal ($m=0$, middle), and prograde ($m=-1$, bottom) modes, for different rotation rates: in blue without rotation, in black with a rotation frequency of 7 $\mu$Hz (0.6 c/d), and in red with a rotation of 15 $\mu$Hz (1.3 c/d). In dotted lines are given the period spacings computed with the TAR.}
    \label{fig:DeltaP_S32}
\end{figure}

\cite{Miglio2008} have conducted an in-depth study of the behaviour of period spacings in high-order g-mode pulsators on the main sequence, i.e. in SPB (Slowly Pulsating B) and $\gamma$ Doradus stars. They have shown theoretically that the period spacings of these stars are not constant if they have a sharp mean molecular weight gradient ($\nabla \mu$) above the convective core, i.e. when they are evolved enough, and do not undergo diffusive mixing at the edge of the core. In that case, the period spacing rather oscillates around a constant value. The mean value depends on the sharpness of the mean molecular weight gradient, whereas the periodicity of that variation, i.e. the number of modes per cycle, is directly related to the location of the sharp feature in $\nabla \mu$.
More recently, \cite{Bouabid2013} explored $\gamma$ Doradus period spacings theoretically, accounting for the effect of uniform rotation in the framework of the TAR. Here we do so, using the non-perturbative approach, and we confirm the main features the authors had identified then. These features are illustrated in Fig. \ref{fig:DeltaP_S32}.

When rotation is included, the period spacings do not oscillate around a constant value, but around a linear trend, that, in the inertial frame of the observer, is decreasing for prograde and zonal modes, and mainly increasing for retrograde modes. As the rotation increases (in Fig. \ref{fig:DeltaP_S32}, from no rotation in blue, to 7 $\mu$Hz, i.e. 0.6 c/d, in black, and then 15 $\mu$Hz, i.e. 1.3 c/d, in red), the slope of the linear trend decreases strongly for prograde modes, less strongly for the zonal modes, and increases for the retrograde ones. In comparison to \cite{Bouabid2013}, these rotation periods correspond to $\Omega / \Omega_c = 0,~ 0.32$, and 0.68, with $\Omega_c = \sqrt{8 G M/27 R^3}$ being the critical breack-up velocity defined by the Roche model.
Additionally, in the inertial frame, the pattern of the variation varies with the period. More precisely, for retrograde modes, the cycle length of the period spacing oscillation increases with period, whereas it decreases for the zonal and retrograde modes. Nevertheless, the number of modes in the pattern, uniquely related to the location of the $\nabla \mu$ \citep[see ][for an elegant demonstration]{Miglio2008}, remains constant.

To sum up, the effect of rotation is two-fold. On the one hand the cycle length of the oscillation in period spacing varies, and on the other hand, the period spacing shows a global linear trend. The slope of this trend depends on the angular degree and azimuthal order, so that: 
\begin{equation}
\Delta \rm P_{n,\ell,m} \, = \, \Sigma_{\ell,m} \, \rm P_{n,\ell,m} + r_{\ell,m},
\end{equation}
where $\Delta \rm P_{n,\ell,m}$ is the period spacing, $\rm P_{n,\ell,m}$ the modes period, and ${\Sigma}_{\ell,m}$ is the slope. 
In the following section (Sect. \ref{Sec_SlopeEffects}), we explore how $\Sigma_{\ell,m}$ varies with different physical parameters, starting with its dependency on the rotation rate.



\subsection{The slope $\Sigma$ as a diagnostic of rotation}
\label{Ssec_slope}
\begin{figure*}
\centering
\hspace{-0.cm}\includegraphics[scale=1]{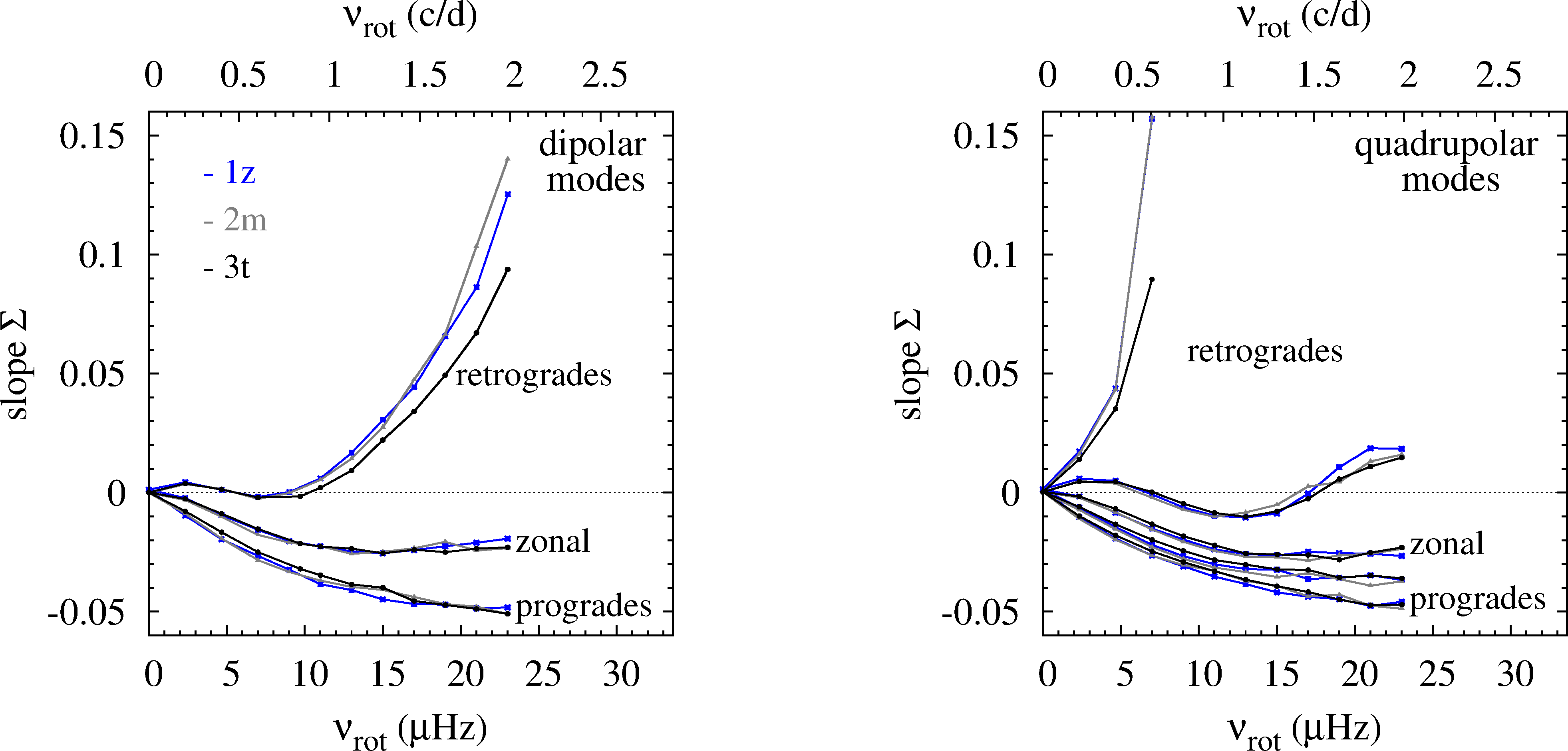}
  \caption{  \label{fig:slopel1-l2}
{\it Left:} Slope of the Period spacing (as a result of a linear fit in a $\Delta P$ versus P diagram) as a function of the stellar model rotation frequency; for dipolar modes ($\ell = 1$) and for models 1z, 2m and 3t which parameters are given in Tab. \ref{tab:models_table}. {\it Right:} same but for quadrupolar modes ($\ell=2$). }
\end{figure*}

In order to explore the dependency of $\Sigma$ on rotation, we performed a series of tests on three different stellar models: 1z, a 1.4 M$_{\odot}$ model on the ZAMS (Zero Age Main Sequence); 2m, a 1.6 M$_{\odot}$ model on the MS; and 3t, a  1.86 M$_{\odot}$ model on the TAMS (Terminal Age Main Sequence, see Table \ref{tab:models_table}), which cover the $\gamma$ Doradus instability strip. For the sake of simplicity, we opted for the models with diffusive transport. The impact of stratification at the edge of the convective core is explored in Sect. \ref{Ssec_mix}. 
For each stellar model, we have computed the pulsation spectra corresponding to uniform rotation frequencies ranging from 0 to 23 $\mu$Hz (i.e. 0.5 to 5 days in period). 

For each pulsation spectrum, we made a linear fit to the period spacing $\Delta P$ as a function of period (such $\Delta P$ versus $P$ diagrams are shown Fig. \ref{fig:DeltaP_S32}). The slope $\Sigma_{\ell,m}$ measured from that linear fit is plotted against the corresponding rotation frequency in Fig. \ref{fig:slopel1-l2} for dipolar (left) and quadrupolar (right) modes .

We see that the behavior of $\Sigma_{\ell,m}$ as a function of the rotation frequency is very similar for the three stellar models that have been explored. Hence, given the differences between those models in terms of internal structure (evolutionary stage, extension of convective and radiative regions, etc..., see Tab. \ref{tab:models_table}), the slope $\Sigma_{\ell,m}$ does not depend on the internal structure. It seems to depend solely on the mode angular degree and azimuthal order $\{ \ell, m \}$, and on the rotation rate of the model.  For the sectorial retrograde modes $\{\ell = 2, m = 2\}$, the slope $\Sigma_{\ell,m}$ has been plotted only up to 11 $\mu$Hz because at higher rotation frequencies, their frequency reaches negative values in the inertial frame, which results in frequencies reflected around zero, therefore obeying a different dynamics \citep[as found also in][]{Bouabid2013}.

The great advantage of $\Sigma$ is that, except for retrograde modes, it does not depend on the radial orders excited and detected, contrary to the mean period spacing. Here we have considered g-modes with radial orders $n$ from -50 to -20, and we have verified that $\Sigma$ does not vary when the radial order range is truncated for prograde and zonal modes. The case of retrogtrade modes is more problematic, Sect. \ref{Ss_retro} is dedicated to this issue. 

To sum up, modes in rapidly rotating $\gamma$ Doradus stars tend to cluster in groups of modes of given $\{ \ell, m \}$, and the period spacings of these groups of modes follow a linear trend with a slope directly related to the rotation rate. As such, the slope $\Sigma_{\ell,m}$ constitutes a promising seismic diagnostic for rotation in $\gamma$ Doradus stars.

\section{Sensitivity of the slope $\Sigma$}
\label{Sec_SlopeEffects}

In the previous section, we have shown that the slope $\Sigma$ of the period spacing as a function of the period seems to depend strongly on the rotation period of the model. We considered three different stellar models of $\gamma$ Doradus stars, in three evolutionary stages of their main sequence. In this section, we further investigate this rotation-slope relation, and the impact of a changing stellar structure on that relation.

\subsection{Sensitivity to the stellar parameters and the metallicity}
\label{Ssect:stelpar}
 \begin{figure*}
\centering
\includegraphics[scale=0.45]{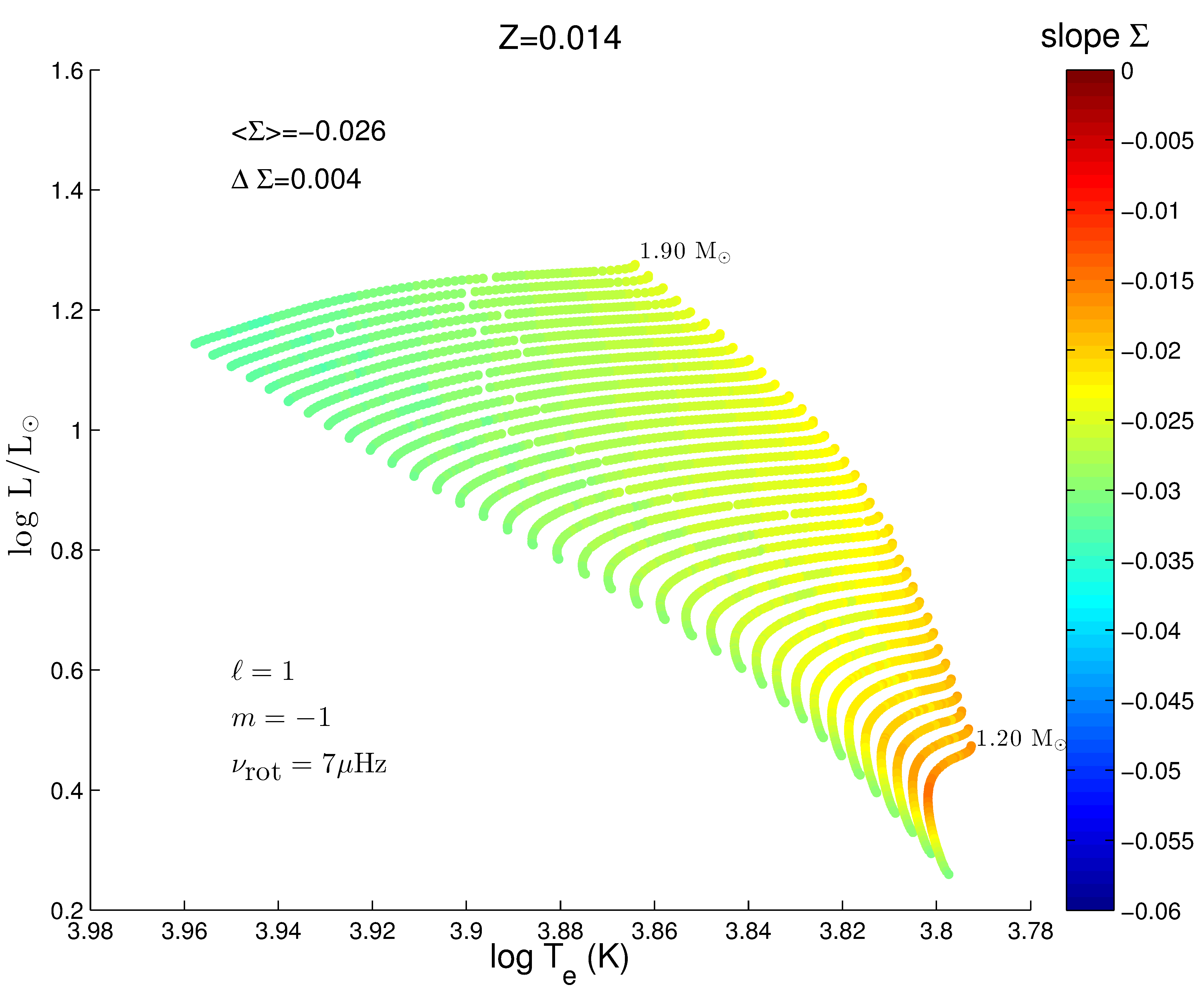}
\hspace*{0.6cm}\includegraphics[scale=0.45]{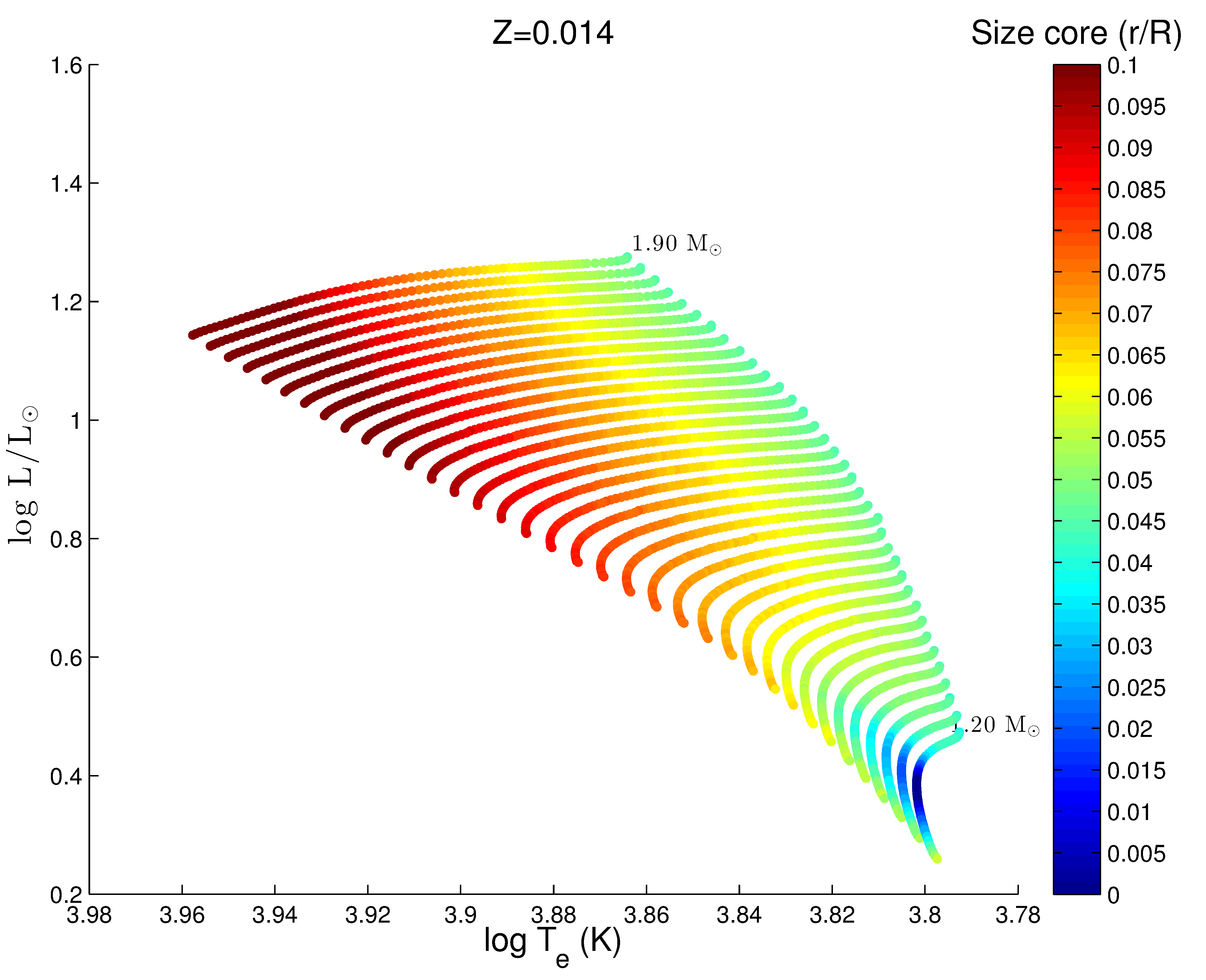}
  \caption{{\it Left:} Slope $\Sigma$ of the period spacing of prograde modes computed with the asymptotic relation (see Eq. \ref{TARAsympt}) in the HR diagram for a grid of models with turbulent diffusion ($D_t = 700 cm^2.s^{-1}$), no overshoot, for masses and $\alpha_{MLT}$ given in Table \ref{tab:models_grid}, for a value of the metallicity of $Z=0.014$ and for a value of the rotation frequency of 7 $\mu$Hz (0.6 c/d). {\it Right:} Size of convective cores in the HR diagram for the same models. \label{fig:conv-core}
}
 \end{figure*}
 \begin{figure}
   \centering
   \includegraphics[scale=0.89]{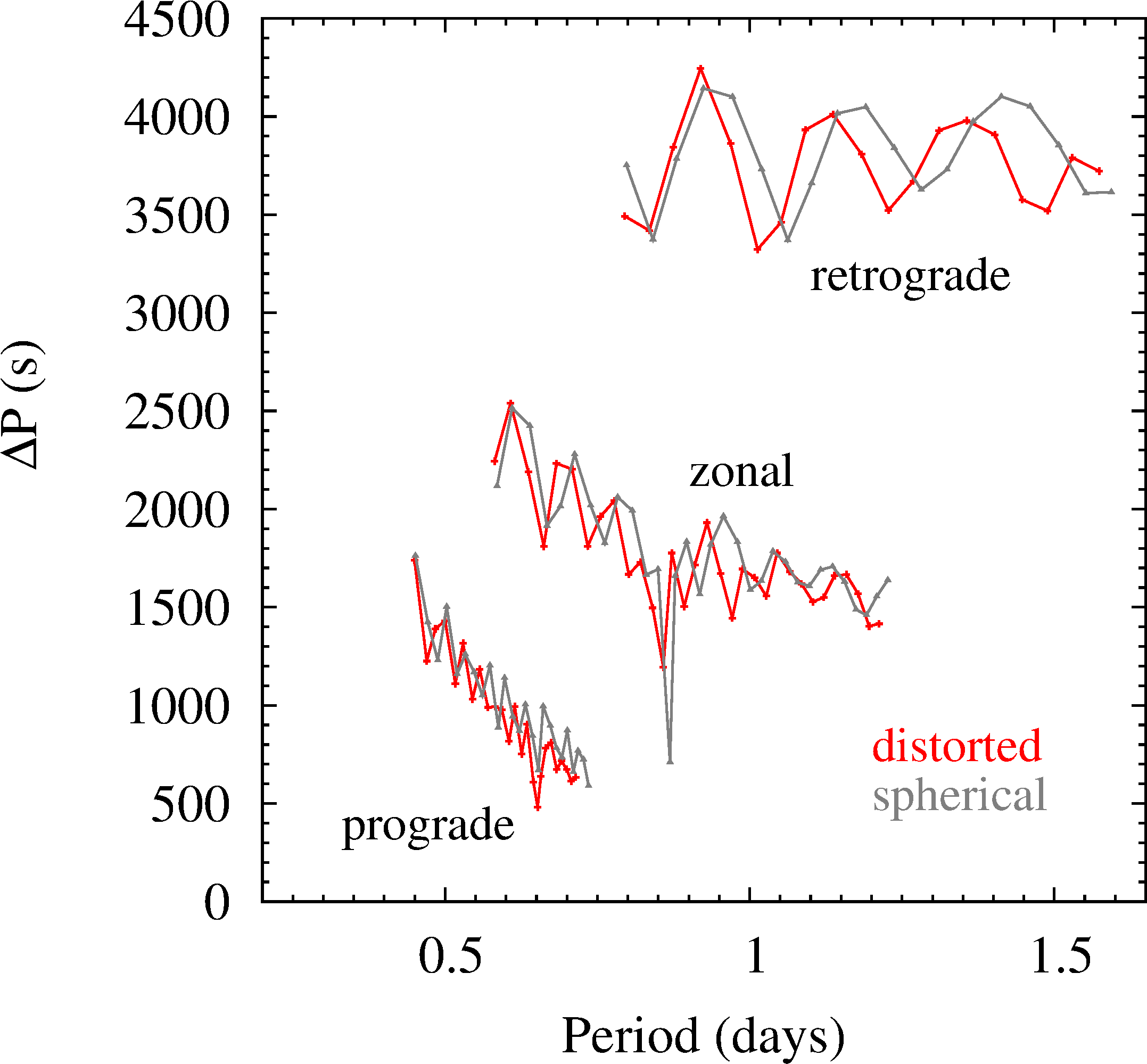}
   \caption{Period spacing as a function of the period for dipolar modes as a function of the period for model 3m' (see Tab. \ref{tab:models_table}). The red curves refer to the distorted model and the grey curves to the corresponding spherical model.}
   \label{fig:dist}
\end{figure}
 In order to study the changes in the slope $\Sigma_{\ell,m}$ for varying stellar parameters, we computed the g-modes spectra for the grid of models given in Table \ref{tab:models_grid}, restricting ourselves to the case with turbulent diffusion and no overshoot, but making use of the asymptotic formulation accounting for rotation. As shown in Sect. \ref{Ss_comparaison_ACOR-TAR}, the validity of the asymptotic formulation is limited. However, it is used here because it allows us to thoroughly explore the $\gamma$ Doradus instability strip which would be numerically significantly more expensive with the TAR or with the complete method. Hence, this approach allows us to assess the variation of the slope when varying the stellar parameters, not its absolute value, which can be erroneous due to the limited validity of the asymptotic method.

\begin{figure*}
	\includegraphics[scale=1]{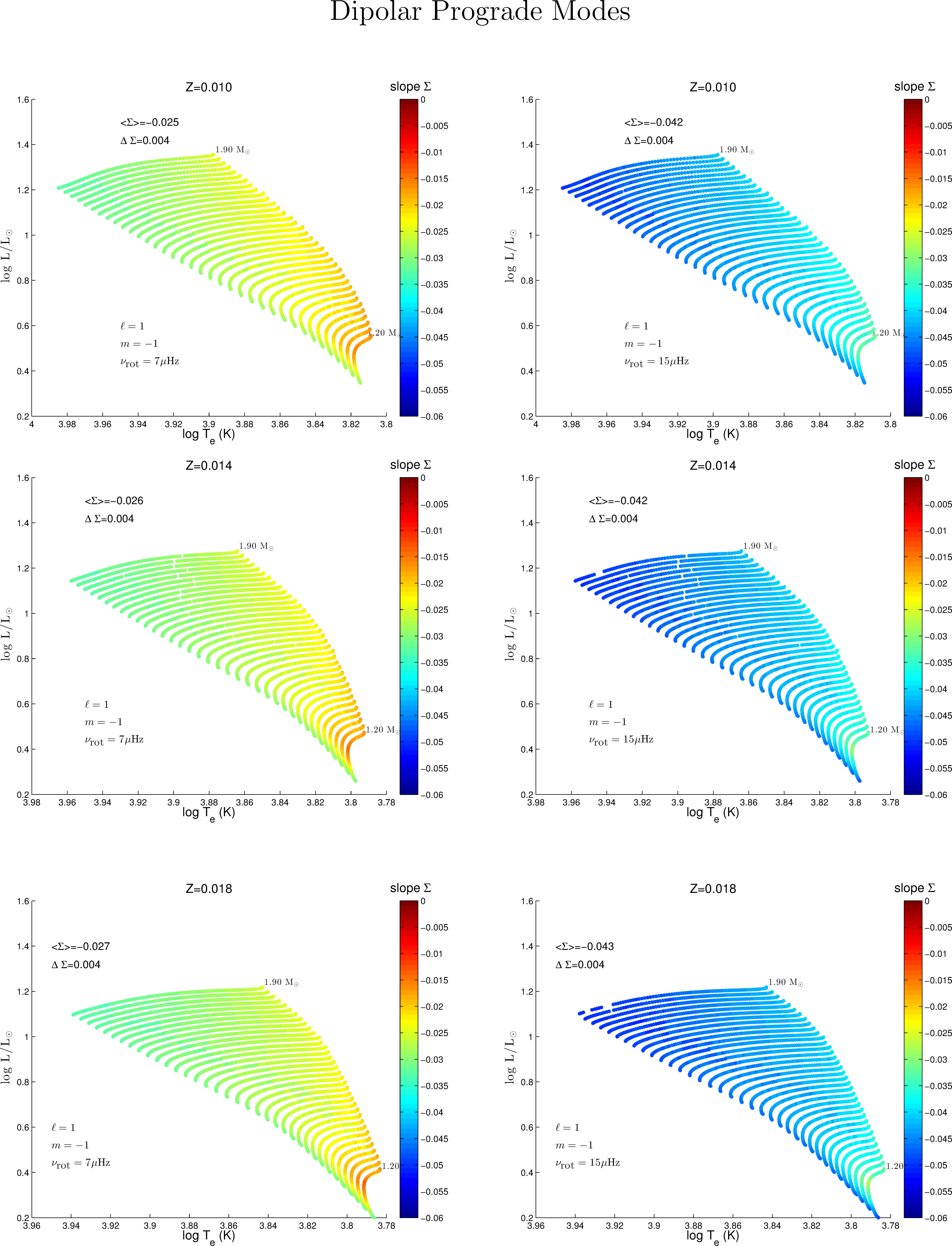}
    \caption{ Slope $\Sigma$ of the period spacing of zonal modes computed with the asymptotic relation (see Eq. \ref{TARAsympt}) in the HR diagram for a grid of models with turbulent diffusion ($D_t = 700 cm^2.s^{-1}$), no overshoot, for masses and $\alpha_{MLT}$ given in Table \ref{tab:models_grid}, for three different value of the metallicity -- from $Z=0.010$ (top) to $Z=0.018$ (bottom)-- and two value of the rotation frequency -- 7 $\mu$Hz (0.6 c/d, left) and 15 $\mu$Hz (1.3 c/d, right).  }
    \label{fig:HR_slope_pro}
\end{figure*}

\begin{figure*}
	\includegraphics[scale=1]{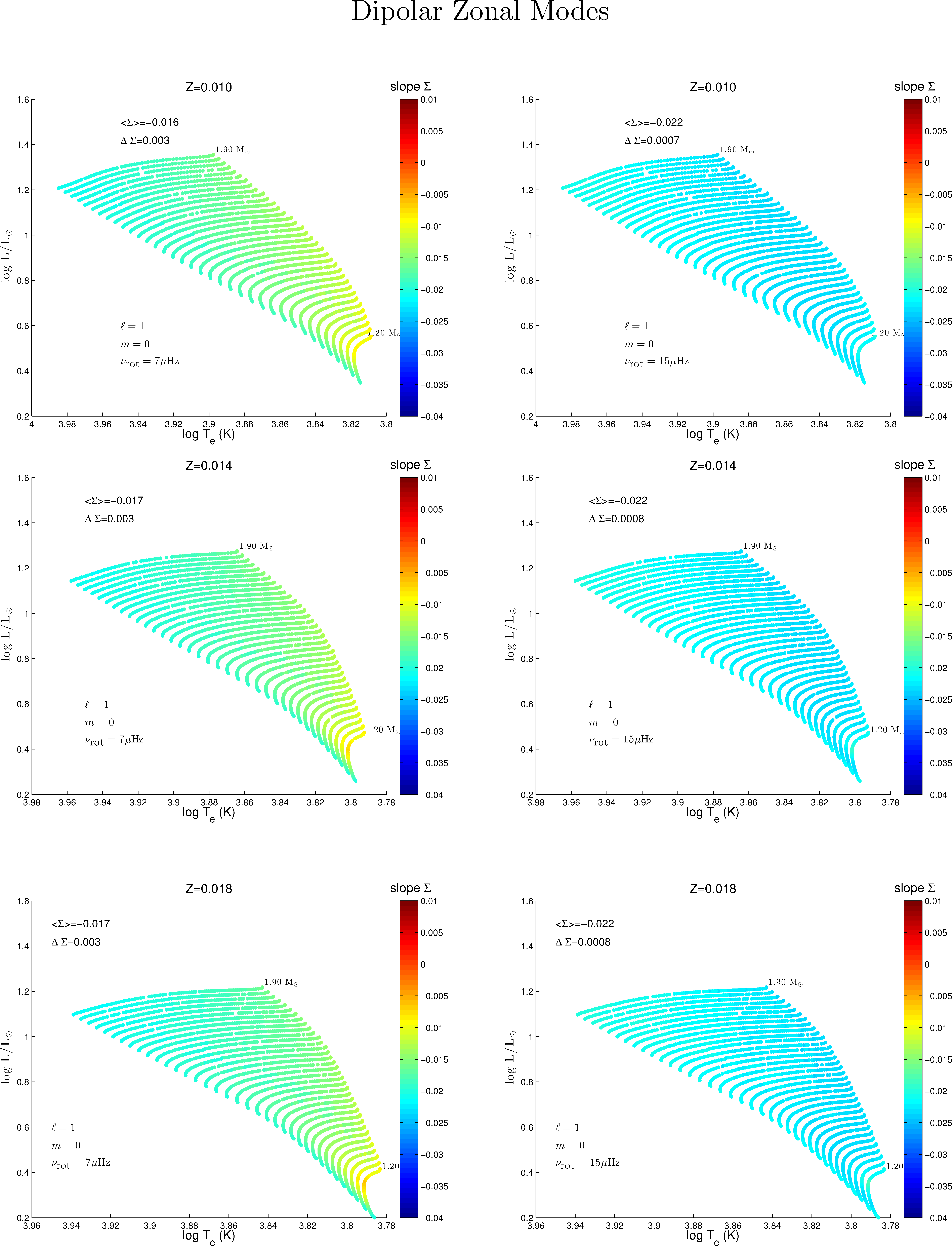}
    \caption{Slope $\Sigma$ of the period spacing of prograde modes computed with the asymptotic relation \ref{TARAsympt} in the HR diagram for a grid of models with turbulent diffusion ($D_t = 700 cm^2.s^{-1}$), no overshoot, for masses and $\alpha_{MLT}$ given in Table \ref{tab:models_grid}, for three different value of the metallicity -- from $Z=0.010$ (top) to $Z=0.018$ (bottom)-- and two value of the rotation frequency --  7 $\mu$Hz (0.6 c/d, left) and 15 $\mu$Hz (1.3 c/d, right). }
    \label{fig:HR_slope_zon}
\end{figure*}

Figure \ref{fig:conv-core} (left) gives the slope $\Sigma$ (in colour scale) for dipolar prograde modes in the grid of models given in Tab. \ref{tab:models_grid}, with turbulent diffusion $D_t=700 \rm cm^2.s^{-1}$ and no overshoot, for a rotation frequency of 7 $\mu$Hz, and a metallicity $Z=0.014$. We refer to Fig. \ref{fig:HR_slope_pro}, Fig. \ref{fig:HR_slope_zon} and Fig. \ref{fig:HR_slope_retro} at the end of the article for a full overview for dipolar modes (prograde, zonal and retrograde modes resp.), for two different rotation frequencies (7 $\mu$Hz and 15 $\mu$Hz), and three different metallicities. Also in these figures, we report on the mean and standard deviation of $\Sigma_{\ell,m}$ at a given rotation period, and for a set of quantum numbers $\{\ell,m\}$.

The first striking feature is the uniformity of the values for the prograde modes, for the zonal modes, as well as for the retrograde modes at the lowest rotation period. For these, mostly the low-mass end of the main sequence in the HR diagram seems to depart significantly from the uniform slope value by two sigma. The low end of the mass range corresponds to the mass limit for the disappearance of the convective core. Therefore, this variation at the low-mass end of the instability strip is due to the drastic differences of internal structure between radiative cores models (with masses ranging approximately from 1.2 to 1.3 M$_{\odot}$) and convective core models everywhere else. To illustrate this point, the size of convective cores in the models covering the instability strip is plotted in Fig. \ref{fig:conv-core} (right). The dark blue points are models that do not manage to sustain a convective core when they leave the ZAMS. This hypothesis is supported by the fact that the slope is closer to the mean value at the beginning of the main sequence, i.e. when the star has developed a convective core, and the discrepancy appears progressively when the core can no longer sustain convection.

The metallicity does not affect the slope in most of the instability strip, except again in the low-mass end (as illustrated in Fig. \ref{fig:HR_slope_pro} and Fig. \ref{fig:HR_slope_zon}). Note that when the metallicity increases, the variation seems to increase and to extend in mass. This is particularly obvious for the dipolar prograde modes, for a rotation of 7 $\mu$Hz (Fig. \ref{fig:HR_slope_pro}, left). This confirms the fact that the origin of the variation is likely due to the abscence of a convective core, given the appearance of a convective core occurs at higher masses with increasing metallicity.

Note that at a given rotation period, this effect is higher for prograde modes than for zonal modes, and in turn higher for zonal modes than for retrograde modes. An exploration of the kernels of modes of different azimuthal orders at a given radial order showed that prograde modes probe deeper layers than zonal and retrograde modes. This confirms the conclusion that the variation is due to the appearance of a convective core, and is all the more important for the modes which probe closer to the core. 

Apart from these local variations in the low-mass end of the HR diagram, the slope $\Sigma$ varies very little with the stellar mass, evolutionary status (along the main sequence) or metallicity. The mean values agree within the standard deviation regardless of the value of the metallicity, and are identical in most cases. The standard deviations themselves are typically one order of magnitude lower than the mean value, and vary from 4\% in the best case (zonal modes at 15 $\mu$Hz of rotation) to 18\% in the worst (for the zonal modes with 7 $\mu$Hz of rotation frequency). The behaviour of retrograde modes will be discussed further in Sect. \ref{Sec_Ccl}. 


\begin{figure}
  \includegraphics[scale=0.89]{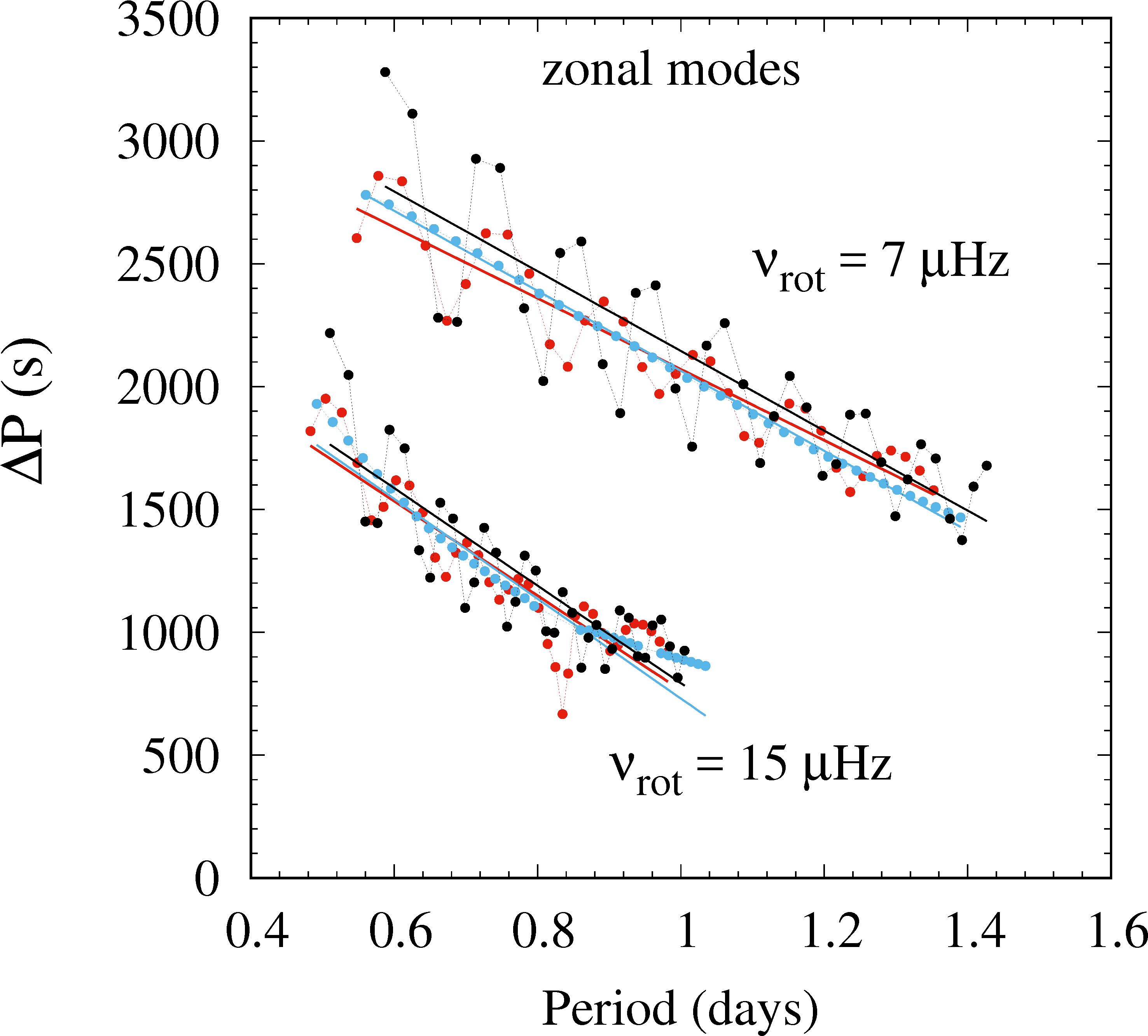}
  \caption{Period spacing as a function of the period for zonal dipolar modes computed for models with turbulent diffusion (3m, blue), overshooting (3m'', black) or none (3m', red, see Tab.\ref{tab:models_table}), with linear fits, for uniform rotations of 7 $\mu$Hz and 15 $\mu$Hz.  }
  \label{fig:mixing_puls}
\end{figure}
\begin{figure}
  \centering
  \includegraphics[scale=0.48]{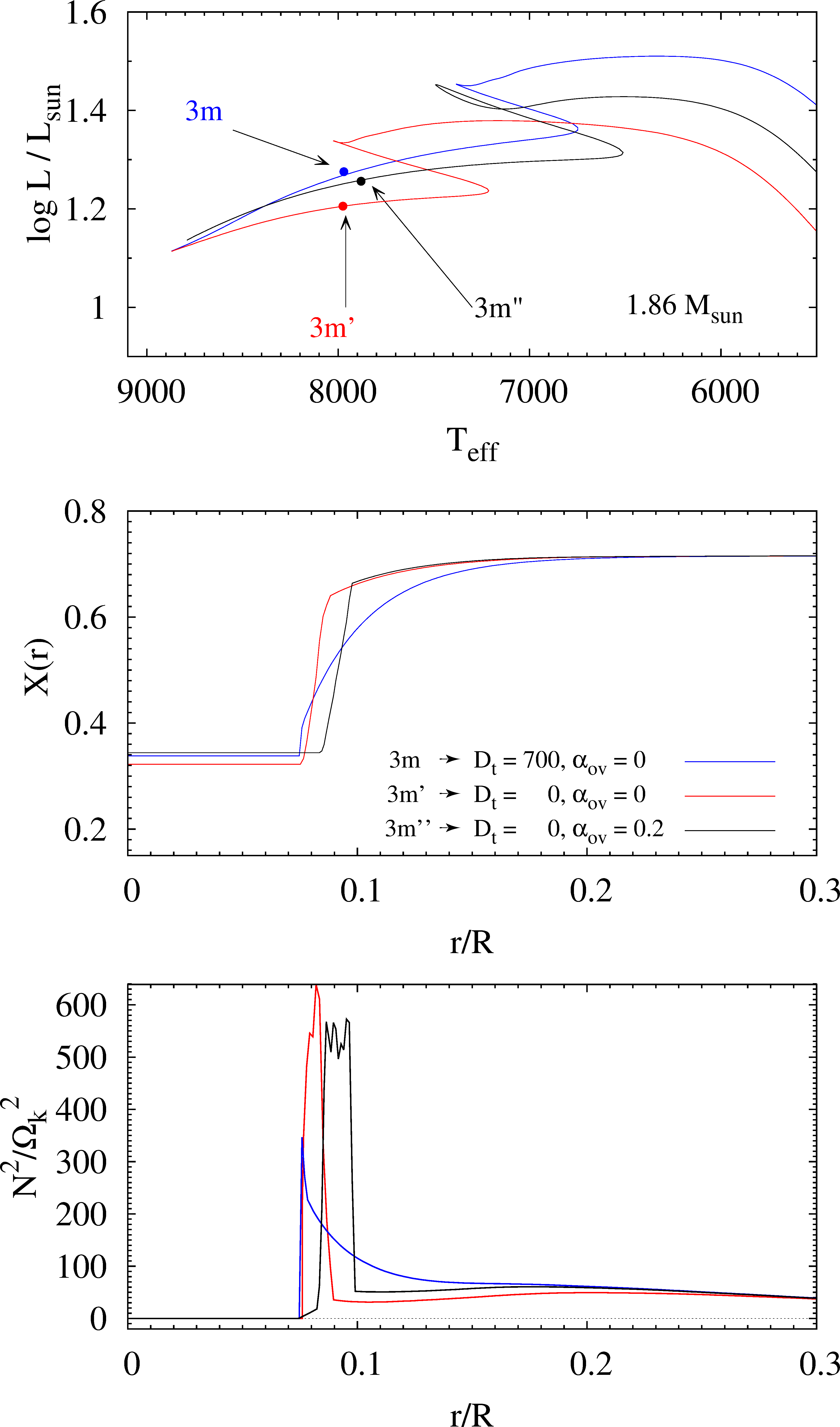}
  \caption{{\it Top:} Hertzsprung-R\"ussell diagram where the three models considered here are shown, together with the evolutionary tracks they belong to. {\it Middle:} Hydrogen profile of the three models. {\it Bottom:} Square of the Brunt-V\"a\"iss\"al\"a frequency scaled by the Keplerian angular velocity, as a function of the fractional radius. The three models are 1.86 M$_{\odot}$, with different stratification at the edge of the convective core: either with turbulent diffusion (3m, blue) or overshooting (3m'', black) or none (3m', red). }
  \label{fig:mixing_model}
\end{figure}

\subsection{Impact of the centrifugal distortion}
\label{Ssec_2D}

\begin{table}
  \centering
  \caption{ Slopes $\Sigma$ of the period spacing ridges for the pulsation spectra of the spherical model ($\Sigma_{\rm 1D}$, left), and of the distorted model ($\Sigma_{\rm 2D}$, right). The uncertainties are in fact spread due to the wiggles in the period spacings, because no mixing was included in these models (see Fig. \ref{fig:dist}).
  }
  \label{tab:slope_1D-2D}
  \begin{tabular}{lcccr} 
		\hline
		modes \, ~ \, & \, ~ \, $\Sigma_{\rm 1D}$ \, ~ \, & \, ~ \, $\Sigma_{\rm 2D}$  \\
                \hline
		prograde \, ~ \, & \, ~ \, -0.035 \, ~ \, & \, ~ \, -0.041 \\
		zonal \, ~ \, & \, ~ \, -0.012 \, ~ \, & \, ~ \, -0.013 \\
		retrograde \, ~ \, & \, ~ \, 0.0010 \, ~ \, & \, ~ \, 0.0009   \\
		\hline
  \end{tabular}
\end{table}

In order to assess the influence of centrifugal distortion on g-mode spectra in $\gamma$ Dor stars, we compared pulsations computed for a spherical model with those computed for an equivalent model, distorted a posteriori. 
The spherical model is labelled 3m' in Table \ref{tab:models_table}, with a differential rotation profile, i.e. a convective core which rotates at around 15 $\mu$Hz (1.3 c/d), and an envelope at a rate of 7 $\mu$Hz (0.6 c/d).
The 2-dimensional model was built following the method developed by \cite{Roxburgh2006}. It consists of building the 2D acoustic structure of a rotating star in hydrostatic equilibrium, starting with spherical profiles of the structural quantities, and including the centrifugal force,  using an iterative scheme. The pulsations of both models were calculated using the ACOR code. For the 2-dimensional model, we used the 2D integration scheme which has been adapted to handle non-barotropic models of stars \citep[for more details, see][]{Ouazzani2015}.
The dipolar modes spectra were computed using 5 spherical harmonics in both schemes. In Fig. \ref{fig:dist} we show the period spacing of dipolar g-modes plotted against the period. Although the periods and the period spacings of individual modes seem to differ significantly, as shown in Fig. \ref{fig:dist}, the global characteristics are conserved: the mean value of the period spacing, the number of modes per pattern, the extent of the pattern in period and, particularly, the slope $\Sigma_{\ell,m}$ for each group of modes of same azimutal order. The slopes obtained in the pulsation spectra for the spherical and the distorted models are give in Tab. \ref{tab:slope_1D-2D}. Note that we do not find higher deviations for the retrograde modes than for the prograde ones. However, the error in slope induced by spherical modelling of the structure is of the same order as the dispersion due to changing stellar parameters or mixing (next section). Additionally, the intense computational resources required by the 2D calculations prevent a global seismic approach at this stage. We will therefore consider spherical modelling for the remainder of this paper.

\subsection{Effect of the diffusive or instantaneous mixing}
\label{Ssec_mix}

To infer the effect of the stratification at the edge of the core, we investigated the g-mode period spacings of three models located close to each other in the HR diagram (see Fig. \ref{fig:mixing_model}, {\it top}), but with three different mixing scenarios.  The model 3m' was computed without any mixing (see Tab.\ref{tab:models_table}, and in black in Fig.\ref{fig:mixing_model}), the model 3m presents diffusive mixing with a coefficient $\rm D_t \,= \,700 cm^{2}.s^{-1}$ (Fig.\ref{fig:mixing_model} in  blue), and model 3m'' includes overshooting over 0.2 local pressure scale heights (Fig.\ref{fig:mixing_model} in red). Note, from the HR diagram (Fig. \ref{fig:mixing_model}, top) that switching on mixing at the convective core boundary, by extending the mixed region, results into a longer core-hydrogen burning phase.  

In Fig. \ref{fig:mixing_model} ({\it Middle}) we give the hydrogen profile of the three models. As illustrated, the effect of turbulent diffusion is to smooth out the composition gradient without changing its position, whereas the convective core overshoot moves the location of the {\it discontinuity}, without impacting the gradient as such.

To understand the effect on pulsations, we plot the Brunt-V\"a\"iss\"al\"a frequency profiles ({\it bottom}), which is the buoyancy frequency under which the g-modes are trapped. Since the mixing changes the location or the sharpness of the gradient of mean molecular weight, it changes the profile of N$^2$, and therefore g-modes cavity.

\begin{figure*}
	\includegraphics[scale=1]{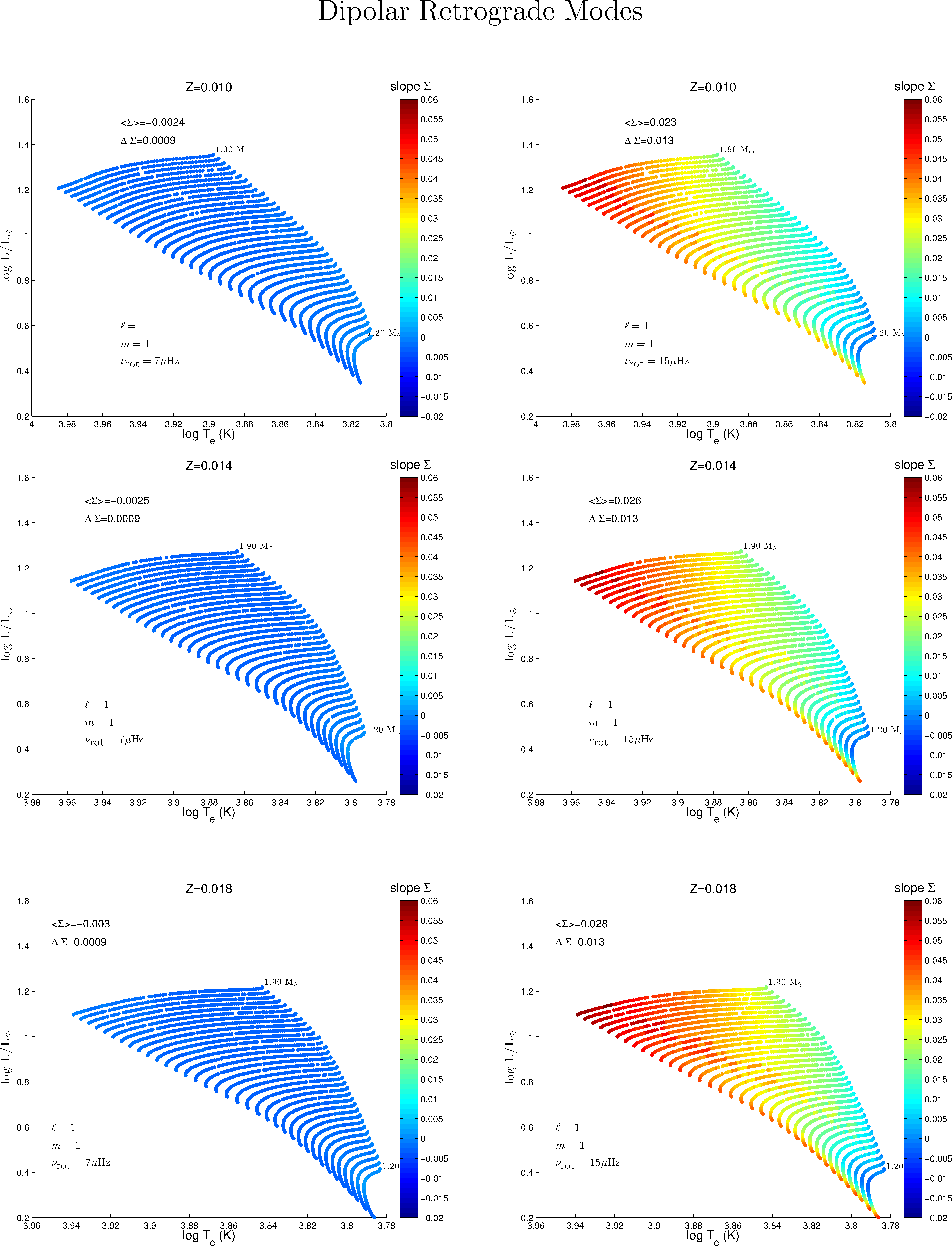}
    \caption{Slope $\Sigma$ of the period spacing of retrograde modes computed with the asymptotic relation \ref{TARAsympt} in the HR diagram for a grid of models with turbulent diffusion ($D_t = 700 cm^2.s^{-1}$), no overshoot, for masses and $\alpha_{MLT}$ given in Table \ref{tab:models_grid}, for three different value of the metallicity-- from $Z=0.010$ (top) to $Z=0.018$ (bottom)-- and two value of the rotation frequency --  7 $\mu$Hz (0.6 c/d, left) and 15 $\mu$Hz (1.3 c/d, right).   }
    \label{fig:HR_slope_retro}
\end{figure*}

The pulsation spectra of these models were computed with the non-perturbative method (Sect. \ref{Ss_ACOR}), for $\ell$=1 modes with radial orders varying from -50 to -20. In Fig. \ref{fig:mixing_puls} is given the period spacing as a function of the period of dipolar zonal modes, for two different rotation frequencies (7$\mu$Hz and 15$\mu$Hz), and for the three mentioned mixing cases. The cases of prograde and retrograde modes are similar in every aspect. As illustrated in Fig. \ref{fig:mixing_puls}, period spacings show various behaviour: a steep molecular-weight gradient (such as in models 3m' and 3m'', red and black resp.), results in a periodic pattern in the period spacing. The periodicity of this behaviour varies with the location of the mixed region boundary \citep{Miglio2008}, the closer the boundary is to the center, the higher the number of modes per pattern. In the presence of diffusive mixing (model 3m, blue), the period spacing shows a smooth behaviour. The aim here is to assess wheteher this has an impact on the value of the slope $\Sigma$. In Tab. \ref{tab:mixing_table} the values of $\Sigma$ for the three different mixing cases and for two rotation periods (corresponding to Fig. \ref{fig:mixing_puls}) are given for dipolar modes. We compare the spread in $\Sigma$ due to different mixing scenarios with the spread due to varying mass and evolutionary stage given in Fig. \ref{fig:HR_slope_pro}, Fig. \ref{fig:HR_slope_zon} and Fig. \ref{fig:HR_slope_retro}. Once again, apart from the retrograde modes (which show a spread of 0.005 in slope), it seems that for the prograde and zonal modes, this spread is of the same order than the dispersion with mass and age (around 0.003). Therefore, we conclude that the mixing at the edge of the convective core does not impact the slope $\Sigma$ of the period spacing of prograde and zonal modes significantly.

\begin{table}
	\centering
	\caption{Slope ${\Sigma}_{\ell,m}$, given by the linear fit of $\Delta P$ versus $P$, for the dipolar modes computed in the three stellar models listed in Tab.\ref{tab:models_table} }
	\label{tab:mixing_table}

        \begin{tabular}{ |l|l|l|l|l| }
          \hline
          ~& Model \# & $m=$-1 & $m=$0 & $m=$1  \\
          \hline
          \multirow{3}{*}{$\nu_{\rm rot} = $ 7.0 $\mu$Hz} &  3m & $-0.030$ & $-0.019$ & $-0.0023$ \\
          & 3m' & $-0.028$ & $-0.016$ & ~\, $0.0003$   \\
          & 3m'' & $-0.030 $ & $-0.019$ & $-0.0007$\\
          \hline
	  \multirow{3}{*}{$\nu_{\rm rot} = $ 15.0 $\mu$Hz} &  3m & $-0.045$ & $-0.024$  & $0.042$ \\
	  & 3m' & $-0.045$ & $-0.022$ & $0.045$ \\
	  & 3m'' & $-0.047$ & $-0.023$ & $0.048$ \\
	  \hline
        \end{tabular}
\end{table}


\section{Results and Discussion}
\label{Sec_Ccl}

In the previous section we have explored the dependency of the parameter $\Sigma$ which is the slope of the period spacing as a function of the period. Here we try to draw a unique relation between this slope and the internal rotation frequency.

\subsection{A one-to-one $\Sigma$ - $ \rm \nu_{rot}$ relation}
\begin{figure}
	\includegraphics[scale=0.72]{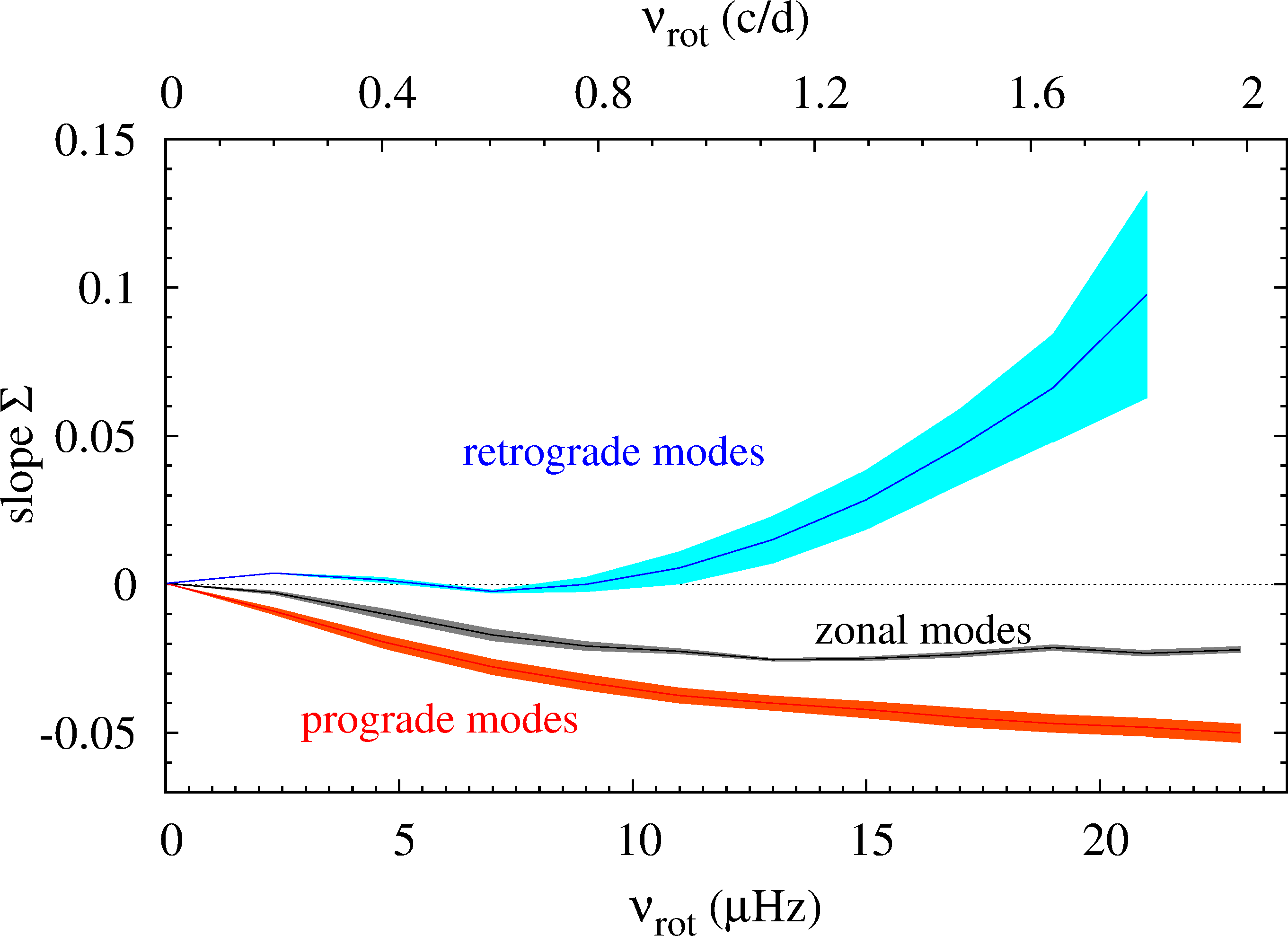}
    \caption{Diagram giving the one-to-one relation between the slope of the period spacing, i.e. the observable $\Sigma$, and the rotation frequency established as an average of the non-perturbative calculations for models 1z, 2m, and 3t. It is given here for dipolar modes: prograde modes in red, zonal modes in black and retrograde modes in blue. The dispersions correspond to the variations of $\Sigma$ due to the mass, age on the main sequence, metallicity, and type of mixing on the edge of the convective core, computed at each rotation rate for the grid of models given in Tab. \ref{tab:models_grid} using the asymptotic formula.}
    \label{fig:diagnostics}
\end{figure}

We evaluate the scatter caused by the differences in structure encountered in the $\gamma$ Dor instability strip. As mentionned in Sect. \ref{Ssect:stelpar}, the asymptotic formulation of the TAR is convenient to determine scatter, even if the absolute values can differ from the non-perturbative calculations. Hence, the strategy is to define the slope-rotation relation by averaging between the values for the models 1z, 2m and 3t. For each rotation frequency, the scatter in slope $\Sigma$ was computed using the asymptotic formula for the grid of models given in Tab. \ref{tab:models_grid}. The result is given in Fig. \ref{fig:diagnostics}, for dipolar modes. This figure shows that there is a one-to-one relation between the slope of the period spacing as a function of the period and the internal rotation frequency in $\gamma$ Doradus stars. This relation shows that the observable $\Sigma$ could be used as a diagnostic for the rotation frequency in the same way that rotational splitting is used in the slowly rotating case.  Note that the relation presented in Fig. \ref{fig:diagnostics} is valid for dipolar modes. Similarly, it is possible to establish a $\Sigma$-rotation relation for quadrupolar modes, as shown in the right pannel of Fig. \ref{fig:slopel1-l2}. 

\subsection{The case of retrograde modes}
\label{Ss_retro}
\begin{figure}
	\includegraphics[scale=0.9]{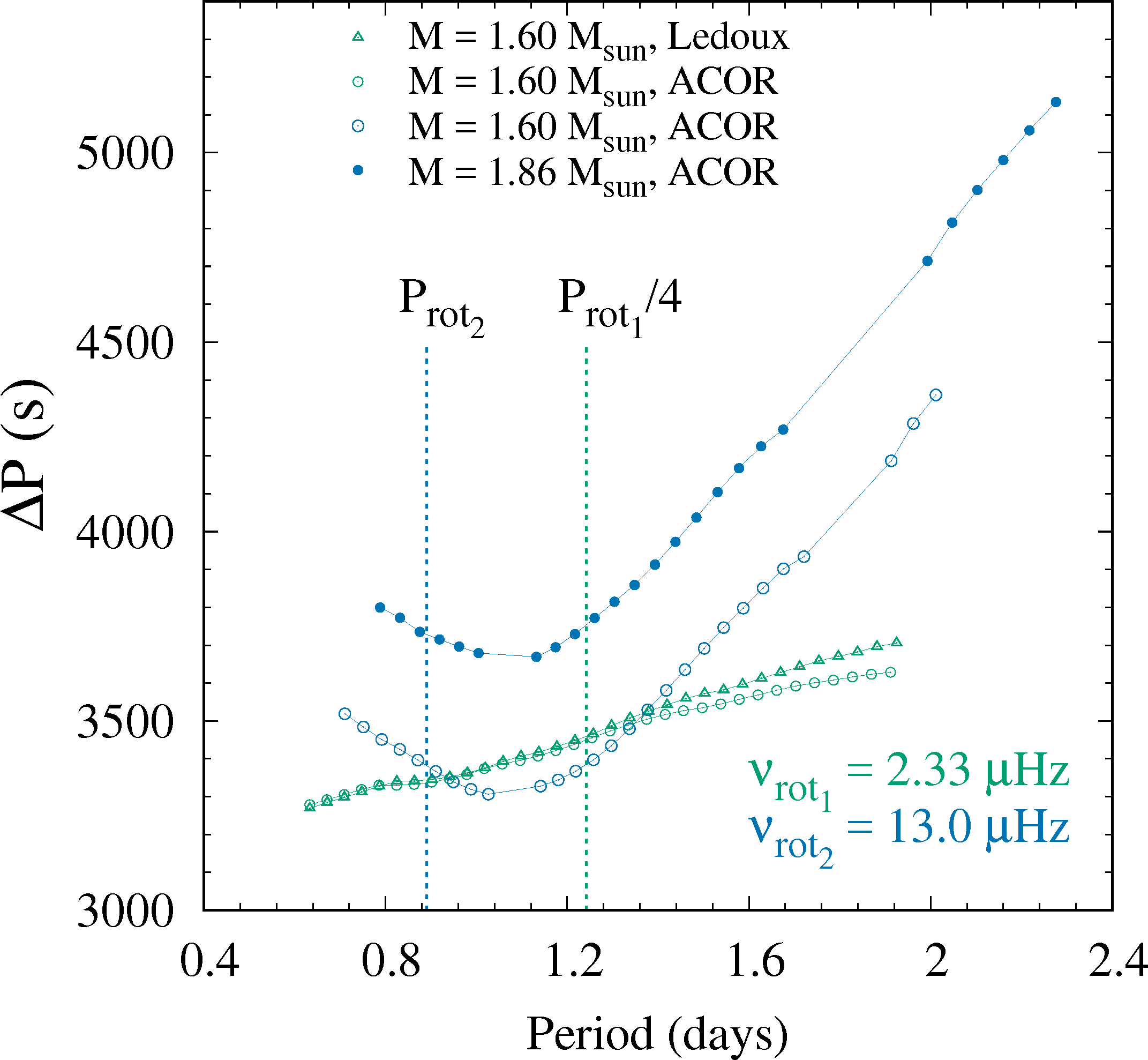}
    \caption{Period spacing as a function of period in the inertial frame for retrograde modes computed for a 1.86 M$_{\odot}$ (filled symbols), and a 1.60 M$_{\odot}$ (open symbols) stellar models, with the ACOR code (circles), or with the Ledoux splitting (triangles), for slow rotation ($\nu_{\rm rot1} = $2.33 $\mu$ Hz, green) and rapid rotation ($\nu_{\rm rot2} = $13.0 $\mu$ Hz, blue). The two vertical dashed lines stand for the period in the inertial frame that equals the rotation period P$_{\rm rot2} = 0.89$ days for the rapid rotation case, and when it equals a quarter of the rotation period for the slowly rotating case P$_{\rm rot1}/4 = 1.24$ days. Some modes are discarded for not presenting a clear $\ell=1$ character.}
    \label{fig:retro}
\end{figure}

Once again, the largest spread is obtained for the retrograde modes. In the inertial frame, the impact of rotation on gravity modes is two fold: (i) the intrinsic effect of rotation on modes dynamics (mainly through the Coriolis force, but also through the centrifugal force), and (ii) the change of reference from the corotating frame to the inertial one. The zonal modes are not affected by the change of reference frame, but prograde and retrograde modes are shifted towards shorter and longer periods, respectively. Depending on the sign of the azimuthal order, this either adds to (for prograde modes) or subtracts from (for retrograde modes) the instrisic effect of rotation. The specificity of retrograde mode period spacing in the inertial frame comes from the competition between the two effects. Although quantitatively inaccurate (see Sect. \ref{Ss_comparaison_ACOR-TAR}), the asymptotic formulation of the period spacing in the traditionnal approximation can help understand this phenomenon qualitatively. From Eq. \ref{TARAsympt}, the period spacing in the inertial frame can be considered as:
\begin{align}
  \Delta {\rm P}_{\rm in}\, \propto \, \frac{1}{\sqrt{\lambda_{\ell,m,s}} \left( 1 - m \frac{{\rm P}_{\rm co}}{{\rm P}_{\rm rot}} \right)}\, ,
  \label{eq:TARAsympt_DPin}
\end{align}
where the factor $\sqrt{\lambda_{\ell,m,s}}\, ^{-1}$ comes from the Coriolis effect on the pulsations in the corotating frame, and $ \left( 1 - m \frac{\rm P_{\rm co}}{\rm P_{\rm rot}} \right)^{-1}$ comes from the change of reference frame. We identify three different regimes for the behavior of retrograde modes period spacing in the inertial frame, illustrated in Fig. \ref{fig:retro}:
\begin{itemize}
\item[-] For {\bf slow rotation}, i.e., when the pulsation periods are significantly smaller than the rotation period, the period spacing follows the behavior given when the effect of rotation is accounted for perturbatively, at first order \citep{Ledoux1951}. This is illustrated in Fig. \ref{fig:retro}, with the two green ridges, calculated for models rotating at 2.33$\mu$Hz ($\rm P_{rot} = 4.97$ days). The green open triangles are the period spacings obtained using the Ledoux formula to derive the retrograde modes periods from the zonal modes ones, whereas the green open circles were obtained with the ACOR non-perturbative calculations. From this Figure, we notice that for inertial periods smaller than $\rm P_{rot}/4$ (straight dashed green line in Fig. \ref{fig:retro}), the inertial period spacing follows the first-order perturbative formalism.
\item[-] For {\bf inertial pulsation periods higher than $\rm \mathbf P_{rot}/4$}, the rotational effect in the inertial frame is then determined by the Eq. \ref{eq:TARAsympt_DPin}. For $\rm P_{in}$ smaller than $\rm P_{rot}$ ($\rm P_{co}$ smaller than $\rm P_{rot}/2$, i.e. in the super-inertial regime), $\rm \Delta P_{in}$ behavior is dominated by the factor $\sqrt{\lambda_{\ell,m,s}}\, ^{-1}$, that is, by the Coriolis impact in the corotating frame. This results in a decreasing contribution to $\rm \Delta P$ with respect to $\rm P$. This is illustrated for the rapidly rotating case in Fig. \ref{fig:retro}, for the parts of the blue ridges that are leftwards of the dashed blue line. Even for the slowly rotating case, this effect is noticeable, as it makes the non-perturbative calculations (green open circles) deviate from the ones based on \cite{Ledoux1951} (green open triangles).      
\item[-] For {\bf $\rm \mathbf P_{in}$ higher than $\rm \mathbf P_{rot}$}, the effect of the change of reference frame dominates. Since $\rm \Delta P_{in}$ goes to infinity when $\rm P_{co} = P_{rot}$, $\rm \Delta P_{in}$ is dominated by an asymptotic behavior towards infinity.
\end{itemize}

In order to explain the spread in $\Sigma$ for retrograde modes shown in Fig. \ref{fig:diagnostics}, we have plotted the retrograde $\rm \Delta P_{in}$ ridges for two different stellar models rotating at the same rate (Fig.\ref{fig:retro}, blue circles). Overall the two blue curves follow the same behavior with respect of the pulsation period, shifted towards higher $\rm \Delta P$ values for the more massive model. The difference between the two ridges resides in the period of modes of given radial order: for the 1.6 M$_{\odot}$ model (blue open circles), the modes are more numerous on the decreasing part of the ridge than for the 1.86  M$_{\odot}$ model (blue filled circles). As a result, when performing a linear fit of these points, the slope of the ridge corresponding to the 1.6 M$_{\odot}$ model is smaller than for the 1.86 M$_{\odot}$ model. In other words, because these ridges are not linear, the periods change of the excited modes (radial orders n between -50 and -15) due to a change of the model's parameters impacts $\Sigma$.
\color{black}{} That explains why the spread in linear slope is much higher for retrograde modes than for zonal or prograde modes, for which the two factors in Eq.\ref{eq:TARAsympt_DPin} go in the same direction. Therefore, should the diagnostic given in Fig. \ref{fig:diagnostics} be used on observations of retrograde modes, we would recommend a more detailed modelling accounting for the period range on which the parameter $\Sigma$ is determined. 

Note that this could be treated more easily in the corotating frame. However, analyses in the inertial frame are needed to relate to seismic observables, which are measured in the inertial frame. In addition, it is relevant to work in the inertial frame in the case of differential rotation, because the change of frame is not straightforward.

\subsection{Application to observations}
\begin{figure}
	\includegraphics[width=0.85\columnwidth]{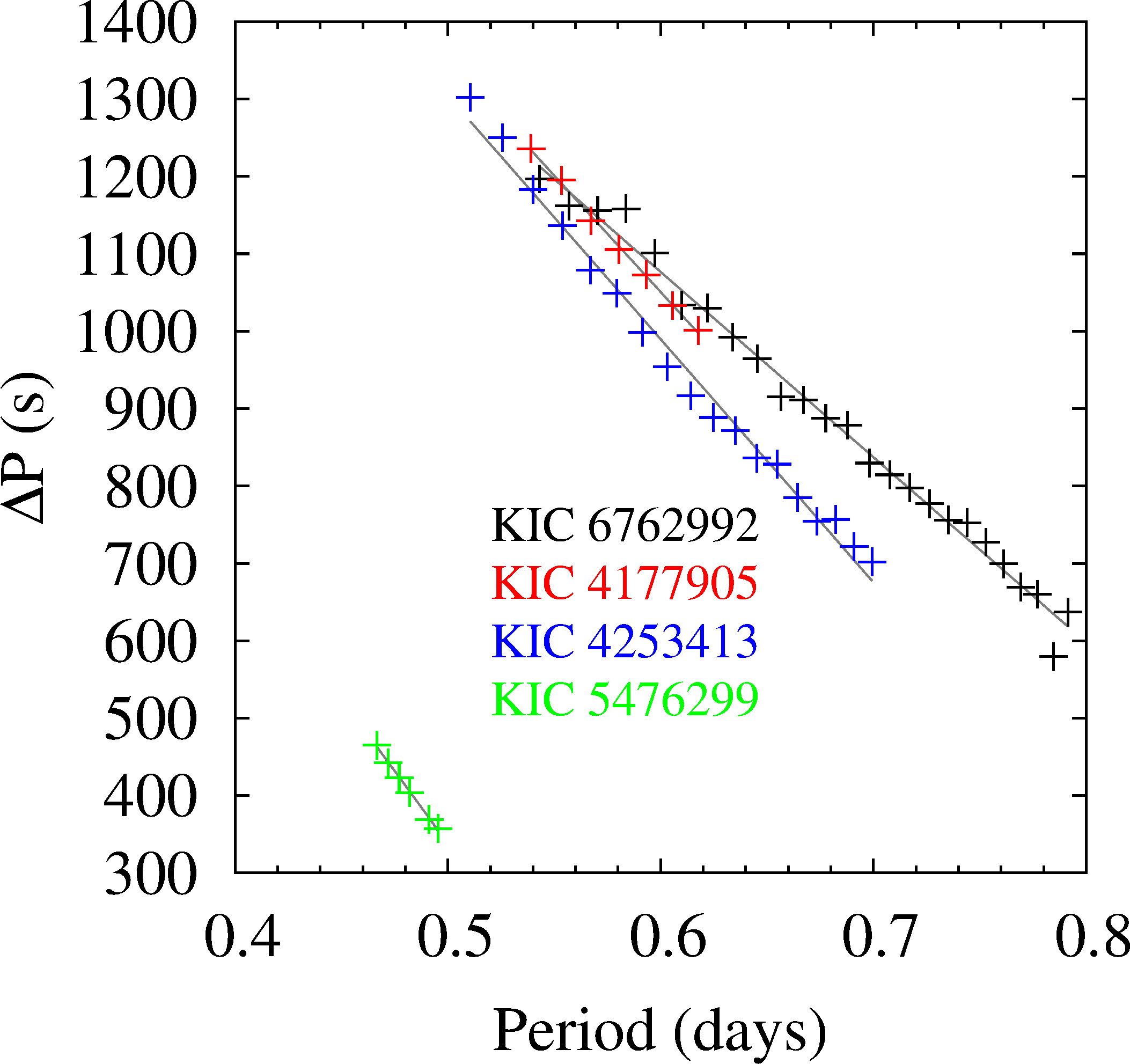}
    \caption{Period spacing as a function of the period for three sequence of modes observed in three stars observed by {\it Kepler}: KIC 6762992 in black points,  KIC 4177905 in red points, and  KIC 4253413 in blue points. The grey lines correpsond to the linear fits used to determine the slope of the respective ridge.}
    \label{fig:keplerstars}
\end{figure}

\begin{table}
  \centering
  \caption{ Results for KIC 6762992, KIC 4177905, KIC 4253413, and  KIC 5476299.
  }
  \label{tab:keplerstars}
  \begin{tabular}{lcccr} 
		\hline
		KIC \, ~ \, & \, ~ \, slope $\Sigma$ \, ~ \, & \, ~ \, $\nu_{\rm rot}$ ($\mu$Hz)  \\
                \hline
		6762992 \, ~ \, & \, ~ \, -0.028 \, ~ \, & \, ~ \, 7.1 $\pm$ 0.9 \\
		4177905 \, ~ \, & \, ~ \, -0.035 \, ~ \, & \, ~ \, 9.8 $\pm$ 1.2 \\
		4253413 \, ~ \, & \, ~ \, -0.036 \, ~ \, & \, ~ \, 10.7 $\pm$ 1.4 \\
		5476299 \, ~ \, & \, ~ \, -0.043 \, ~ \, & \, ~ \, 17.8 $\pm$ 2.9 \\
		\hline
  \end{tabular}
\end{table}

As a proof of concept, we use four stars observed by Kepler, one of which (KIC 4253413) was previously presented in \cite{Bedding2015}. These stars have been observed quasi-continuously by the {\it Kepler} spacecraft for 18 (KIC 4253413, and KIC 6762992), 17 quarters (KIC 5476299), and 15 quarters (KIC 4177905) respectively. The measured oscillation periods shown in the period spacing versus period diagram Fig. \ref{fig:keplerstars} were extracted through the classical pre-whitening procedure using Period04 \citep{Lenz2005}. 

In order to apply the diagnostics based on the slope of the period spacing sequence, $\Sigma$, it is necessary to identify the angular degree and azimuthal order associated to an observed sequence of g-modes. To do so, we rely on the combined knowledge of the mean value of the period spacing, of the slope $\Sigma$, and of the range of observed periods. The mean value of the period spacing mainly carries information on the angular degree $\ell$. As seen in Fig. \ref{fig:DeltaPm_S33}, it can be drastically changed by rotation, therefore a firm identification requires the knowledge of the other observables. As seen in Fig. \ref{fig:diagnostics}, when negative $\Sigma$ is found, there is a need to distinguish between zonal and prograde modes. An easy case arises when we deal with dipolar modes, with steep slopes (i.e. $\Sigma$ values lower than -0.025), because only prograde modes can reach these values. In that case, one should rely on the range of observed modes. The shift in frequency prograde and retrograde modes undergo depends on rotation, and should be consistent with the value of $\Sigma$.    

Hence, based on the range of periods, the period spacings in the range of observed periods, and the slope of the period spacing, the three sequences of modes could unambiguously be identified as dipolar prograde modes. Indeed, the slopes for all the stars are too steep to be attributed to zonal or retrograde modes, and the mean value of the period spacings are not compatible with modes of higher degree $\ell$. The slopes of the period spacing as a function of the period spacing was determined using a simple linear fit of the data points. The diagram given in Fig. \ref{fig:diagnostics} was then used in order to determine the rotation frequency as a well as the respective uncertainty. The latter is due to our lack of additional constraint on the star's structure, i.e. stellar parameters and internal physics. A refined approach with tailored modelling in a star by star approach would help reducing this uncertainty, but this is out of the scope of this paper.

\subsection{Conclusion}
Based on the non uniformity of the period spacings of $\gamma$ Doradus stars, we defined a new observable $\Sigma$, the slope of the period spacing when plotted as a function of period. After a series of tests, we showed that this observable is uniquely related to the internal rotation. In this study, we have assumed a uniform rotation, the case of differential rotation will be investigated in a forthcoming paper. The relation between $\Sigma$ and the internal rotation applies widely on the main-sequence part of the instability strip of $\gamma$ Doradus stars. As a proof of concept, we apply the new asteroseismic diagnostic to a handful of $\gamma$ Doradus stars observed by {\it Kepler}. Using the g-modes observed in these stars, we are able to measure the internal rotation on the lower main sequence, which is still not possible in Sun-like stars. Combined with the core rotation rates determined in red giant stars \citep[see][and references therein]{Mosser2012b}, these measurements will provide stringent contraints on angular momentum transport along the evolution of low-mass stars. The method presented here will allow one to retrieve internal rotation rates for a large sample of stars, which is particularly relevant in the context of the upcoming space missions TESS, and PLATO. 


\section*{Acknowledgements}
 The authors would like to thank Marc-Antoine Dupret and Josefina Montalb\'an for fruitful discussions, and J\o{}rgen Christensen-Dalsgaard for careful reading. Funding for the Stellar Astrophysics Centre is provided by The Danish National Research Foundation. The research is supported by the ASTERISK project (ASTERoseismic Investigations with SONG and Kepler) funded by the European Research Council (Grant agreement no.: 267864).





\bibliographystyle{mnras}


\begin{thebibliography}{}
\makeatletter
\relax
\def\mn@urlcharsother{\let\do\@makeother \do\$\do\&\do\#\do\^\do\_\do\%\do\~}
\def\mn@doi{\begingroup\mn@urlcharsother \@ifnextchar [ {\mn@doi@}
  {\mn@doi@[]}}
\def\mn@doi@[#1]#2{\def\@tempa{#1}\ifx\@tempa\@empty \href
  {http://dx.doi.org/#2} {doi:#2}\else \href {http://dx.doi.org/#2} {#1}\fi
  \endgroup}
\def\mn@eprint#1#2{\mn@eprint@#1:#2::\@nil}
\def\mn@eprint@arXiv#1{\href {http://arxiv.org/abs/#1} {{\tt arXiv:#1}}}
\def\mn@eprint@dblp#1{\href {http://dblp.uni-trier.de/rec/bibtex/#1.xml}
  {dblp:#1}}
\def\mn@eprint@#1:#2:#3:#4\@nil{\def\@tempa {#1}\def\@tempb {#2}\def\@tempc
  {#3}\ifx \@tempc \@empty \let \@tempc \@tempb \let \@tempb \@tempa \fi \ifx
  \@tempb \@empty \def\@tempb {arXiv}\fi \@ifundefined
  {mn@eprint@\@tempb}{\@tempb:\@tempc}{\expandafter \expandafter \csname
  mn@eprint@\@tempb\endcsname \expandafter{\@tempc}}}

\bibitem[\protect\citeauthoryear{{Abt} \& {Morrell}}{{Abt} \&
  {Morrell}}{1995}]{Abt1995}
{Abt} H.~A.,  {Morrell} N.~I.,  1995, \mn@doi [\apjs] {10.1086/192182}, \href
  {http://adsabs.harvard.edu/abs/1995ApJS...99..135A} {99, 135}

\bibitem[\protect\citeauthoryear{{Angulo} et~al.,}{{Angulo}
  et~al.}{1999}]{Angulo1999}
{Angulo} C.,  et~al., 1999, \mn@doi [Nuclear Physics A]
  {10.1016/S0375-9474(99)00030-5}, \href
  {http://cdsads.u-strasbg.fr/abs/1999NuPhA.656....3A} {656, 3}

\bibitem[\protect\citeauthoryear{{Asplund}, {Grevesse}, {Sauval}  \&
  {Scott}}{{Asplund} et~al.}{2009}]{Asplund2009}
{Asplund} M.,  {Grevesse} N.,  {Sauval} A.~J.,   {Scott} P.,  2009, \mn@doi
  [\araa] {10.1146/annurev.astro.46.060407.145222}, \href
  {http://adsabs.harvard.edu/abs/2009ARA26A..47..481A} {47, 481}

\bibitem[\protect\citeauthoryear{{Ballot}, {Ligni{\`e}res}, {Prat}, {Reese}  \&
  {Rieutord}}{{Ballot} et~al.}{2012}]{Ballot2012}
{Ballot} J.,  {Ligni{\`e}res} F.,  {Prat} V.,  {Reese} D.~R.,   {Rieutord} M.,
  2012, in {Shibahashi} H.,  {Takata} M.,   {Lynas-Gray} A.~E.,  eds,
  Astronomical Society of the Pacific Conference Series Vol. 462, Progress in
  Solar/Stellar Physics with Helio- and Asteroseismology. p.~389

\bibitem[\protect\citeauthoryear{{Beck} et~al.,}{{Beck}
  et~al.}{2012}]{Beck2012}
{Beck} P.~G.,  et~al., 2012, \mn@doi [\nat] {10.1038/nature10612}, \href
  {http://adsabs.harvard.edu/abs/2012Natur.481...55B} {481, 55}

\bibitem[\protect\citeauthoryear{{Bedding}, {Murphy}, {Colman}  \&
  {Kurtz}}{{Bedding} et~al.}{2015}]{Bedding2015}
{Bedding} T.~R.,  {Murphy} S.~J.,  {Colman} I.~L.,   {Kurtz} D.~W.,  2015, in
  European Physical Journal Web of Conferences. p.~1005 (\mn@eprint {arXiv}
  {1411.1883}), \mn@doi{10.1051/epjconf/201510101005}

\bibitem[\protect\citeauthoryear{{Belkacem} et~al.,}{{Belkacem}
  et~al.}{2015}]{Belkacem2015b}
{Belkacem} K.,  et~al., 2015, \mn@doi [\aap] {10.1051/0004-6361/201526043},
  \href {http://adsabs.harvard.edu/abs/2015A%26A...579A..31B} {579, A31}

\bibitem[\protect\citeauthoryear{{Berthomieu}, {Gonczi}, {Graff}, {Provost}  \&
  {Rocca}}{{Berthomieu} et~al.}{1978}]{Berthomieu1978}
{Berthomieu} G.,  {Gonczi} G.,  {Graff} P.,  {Provost} J.,   {Rocca} A.,  1978,
  \aap, \href {http://adsabs.harvard.edu/abs/1978A%26A....70..597B} {70, 597}

\bibitem[\protect\citeauthoryear{{B{\"o}hm-Vitense}}{{B{\"o}hm-Vitense}}{1958}]{Bohm-Vitense1958}
{B{\"o}hm-Vitense} E.,  1958, \zap, \href
  {http://cdsads.u-strasbg.fr/abs/1958ZA.....46..108B} {46, 108}

\bibitem[\protect\citeauthoryear{{Bouabid}, {Dupret}, {Salmon},
  {Montalb{\'a}n}, {Miglio}  \& {Noels}}{{Bouabid} et~al.}{2013}]{Bouabid2013}
{Bouabid} M.-P.,  {Dupret} M.-A.,  {Salmon} S.,  {Montalb{\'a}n} J.,  {Miglio}
  A.,   {Noels} A.,  2013, \mn@doi [\mnras] {10.1093/mnras/sts517}, \href
  {http://cdsads.u-strasbg.fr/abs/2013MNRAS.429.2500B} {429, 2500}

\bibitem[\protect\citeauthoryear{{Cameron} et~al.,}{{Cameron}
  et~al.}{2008}]{Cameron2008}
{Cameron} C.,  et~al., 2008, \mn@doi [\apj] {10.1086/590369}, \href
  {http://adsabs.harvard.edu/abs/2008ApJ...685..489C} {685, 489}

\bibitem[\protect\citeauthoryear{{Cantiello}, {Mankovich}, {Bildsten},
  {Christensen-Dalsgaard}  \& {Paxton}}{{Cantiello}
  et~al.}{2014}]{Cantiello2014}
{Cantiello} M.,  {Mankovich} C.,  {Bildsten} L.,  {Christensen-Dalsgaard} J.,
  {Paxton} B.,  2014, \mn@doi [\apj] {10.1088/0004-637X/788/1/93}, \href
  {http://adsabs.harvard.edu/abs/2014ApJ...788...93C} {788, 93}

\bibitem[\protect\citeauthoryear{{Ceillier}, {Eggenberger}, {Garc{\'{\i}}a}  \&
  {Mathis}}{{Ceillier} et~al.}{2013}]{Ceillier2013}
{Ceillier} T.,  {Eggenberger} P.,  {Garc{\'{\i}}a} R.~A.,   {Mathis} S.,  2013,
  \mn@doi [\aap] {10.1051/0004-6361/201321473}, \href
  {http://adsabs.harvard.edu/abs/2013A%26A...555A..54C} {555, A54}

\bibitem[\protect\citeauthoryear{{Cowling}}{{Cowling}}{1941}]{Cowling1941}
{Cowling} T.~G.,  1941, \mnras, \href
  {http://cdsads.u-strasbg.fr/abs/1941MNRAS.101..367C} {101, 367}

\bibitem[\protect\citeauthoryear{{Deheuvels} et~al.,}{{Deheuvels}
  et~al.}{2012}]{Deheuvels2012}
{Deheuvels} S.,  et~al., 2012, \mn@doi [\apj] {10.1088/0004-637X/756/1/19},
  \href {http://adsabs.harvard.edu/abs/2012ApJ...756...19D} {756, 19}

\bibitem[\protect\citeauthoryear{{Dintrans} \& {Rieutord}}{{Dintrans} \&
  {Rieutord}}{2000}]{Dintrans2000}
{Dintrans} B.,  {Rieutord} M.,  2000, \aap, \href
  {http://cdsads.u-strasbg.fr/abs/2000A%26A...354...86D} {354, 86}

\bibitem[\protect\citeauthoryear{{Dupret}, {Grigahc{\`e}ne}, {Garrido},
  {Gabriel}  \& {Scuflaire}}{{Dupret} et~al.}{2005}]{Dupret2005}
{Dupret} M.-A.,  {Grigahc{\`e}ne} A.,  {Garrido} R.,  {Gabriel} M.,
  {Scuflaire} R.,  2005, \mn@doi [\aap] {10.1051/0004-6361:20041817}, \href
  {http://adsabs.harvard.edu/abs/2005A%26A...435..927D} {435, 927}

\bibitem[\protect\citeauthoryear{{Eckart}}{{Eckart}}{1960}]{Eckart1960}
{Eckart} C.,  1960, \mn@doi [Quarterly Journal of the Royal Meteorological
  Society] {10.1002/qj.49708938224}, 89, 567

\bibitem[\protect\citeauthoryear{{Eggenberger}, {Montalb{\'a}n}  \&
  {Miglio}}{{Eggenberger} et~al.}{2012}]{Eggenberger2012}
{Eggenberger} P.,  {Montalb{\'a}n} J.,   {Miglio} A.,  2012, \mn@doi [\aap]
  {10.1051/0004-6361/201219729}, \href
  {http://adsabs.harvard.edu/abs/2012A%26A...544L...4E} {544, L4}

\bibitem[\protect\citeauthoryear{{Ferguson}, {Alexander}, {Allard}, {Barman},
  {Bodnarik}, {Hauschildt}, {Heffner-Wong}  \& {Tamanai}}{{Ferguson}
  et~al.}{2005}]{Ferguson2005}
{Ferguson} J.~W.,  {Alexander} D.~R.,  {Allard} F.,  {Barman} T.,  {Bodnarik}
  J.~G.,  {Hauschildt} P.~H.,  {Heffner-Wong} A.,   {Tamanai} A.,  2005,
  \mn@doi [\apj] {10.1086/428642}, \href
  {http://cdsads.u-strasbg.fr/abs/2005ApJ...623..585F} {623, 585}

\bibitem[\protect\citeauthoryear{{Formicola} et~al.,}{{Formicola}
  et~al.}{2004}]{Formicola2004}
{Formicola} A.,  et~al., 2004, \mn@doi [Physics Letters B]
  {10.1016/j.physletb.2004.03.092}, \href
  {http://cdsads.u-strasbg.fr/abs/2004PhLB..591...61F} {591, 61}

\bibitem[\protect\citeauthoryear{{Fuller}, {Lecoanet}, {Cantiello}  \&
  {Brown}}{{Fuller} et~al.}{2014}]{Fuller2014}
{Fuller} J.,  {Lecoanet} D.,  {Cantiello} M.,   {Brown} B.,  2014, \mn@doi
  [\apj] {10.1088/0004-637X/796/1/17}, \href
  {http://adsabs.harvard.edu/abs/2014ApJ...796...17F} {796, 17}

\bibitem[\protect\citeauthoryear{{Godart}, {Noels}, {Dupret}  \&
  {Lebreton}}{{Godart} et~al.}{2009}]{Godart2009}
{Godart} M.,  {Noels} A.,  {Dupret} M.-A.,   {Lebreton} Y.,  2009, \mn@doi
  [\mnras] {10.1111/j.1365-2966.2009.14903.x}, \href
  {http://adsabs.harvard.edu/abs/2009MNRAS.396.1833G} {396, 1833}

\bibitem[\protect\citeauthoryear{{Grigahc{\`e}ne} et~al.,}{{Grigahc{\`e}ne}
  et~al.}{2010}]{Grigahcene2010}
{Grigahc{\`e}ne} A.,  et~al., 2010, \mn@doi [\apjl]
  {10.1088/2041-8205/713/2/L192}, \href
  {http://adsabs.harvard.edu/abs/2010ApJ...713L.192G} {713, L192}

\bibitem[\protect\citeauthoryear{{Guzik}, {Kaye}, {Bradley}, {Cox}  \&
  {Neuforge}}{{Guzik} et~al.}{2000}]{Guzik2000}
{Guzik} J.~A.,  {Kaye} A.~B.,  {Bradley} P.~A.,  {Cox} A.~N.,   {Neuforge} C.,
  2000, \mn@doi [\apjl] {10.1086/312908}, \href
  {http://adsabs.harvard.edu/abs/2000ApJ...542L..57G} {542, L57}

\bibitem[\protect\citeauthoryear{{Hareter}}{{Hareter}}{2012}]{Hareter2012}
{Hareter} M.,  2012, \mn@doi [Astronomische Nachrichten]
  {10.1002/asna.201211790}, \href
  {http://adsabs.harvard.edu/abs/2012AN....333.1048H} {333, 1048}

\bibitem[\protect\citeauthoryear{{Heger}, {Woosley}  \& {Spruit}}{{Heger}
  et~al.}{2005}]{Heger2005}
{Heger} A.,  {Woosley} S.~E.,   {Spruit} H.~C.,  2005, \mn@doi [\apj]
  {10.1086/429868}, \href {http://adsabs.harvard.edu/abs/2005ApJ...626..350H}
  {626, 350}

\bibitem[\protect\citeauthoryear{{Iglesias} \& {Rogers}}{{Iglesias} \&
  {Rogers}}{1996}]{Iglesias1996}
{Iglesias} C.~A.,  {Rogers} F.~J.,  1996, \mn@doi [\apj] {10.1086/177381},
  \href {http://cdsads.u-strasbg.fr/abs/1996ApJ...464..943I} {464, 943}

\bibitem[\protect\citeauthoryear{{Keen}, {Bedding}, {Murphy}, {Schmid},
  {Aerts}, {Tkachenko}, {Ouazzani}  \& {Kurtz}}{{Keen} et~al.}{2015}]{Keen2015}
{Keen} M.~A.,  {Bedding} T.~R.,  {Murphy} S.~J.,  {Schmid} V.~S.,  {Aerts} C.,
  {Tkachenko} A.,  {Ouazzani} R.-M.,   {Kurtz} D.~W.,  2015, \mn@doi [\mnras]
  {10.1093/mnras/stv2107}, \href
  {http://cdsads.u-strasbg.fr/abs/2015MNRAS.454.1792K} {454, 1792}

\bibitem[\protect\citeauthoryear{{Kippenhahn} \& {Weigert}}{{Kippenhahn} \&
  {Weigert}}{1994}]{Kippenhahn1994}
{Kippenhahn} R.,  {Weigert} A.,  1994, {Stellar Structure and Evolution}

\bibitem[\protect\citeauthoryear{{Kurtz}, {Saio}, {Takata}, {Shibahashi},
  {Murphy}  \& {Sekii}}{{Kurtz} et~al.}{2014}]{Kurtz2014}
{Kurtz} D.~W.,  {Saio} H.,  {Takata} M.,  {Shibahashi} H.,  {Murphy} S.~J.,
  {Sekii} T.,  2014, \mn@doi [\mnras] {10.1093/mnras/stu1329}, \href
  {http://adsabs.harvard.edu/abs/2014MNRAS.444..102K} {444, 102}

\bibitem[\protect\citeauthoryear{{Kurucz}}{{Kurucz}}{1998}]{Kurucz1998}
{Kurucz} R.~L.,  1998, Highlights of Astronomy, \href
  {http://cdsads.u-strasbg.fr/abs/1998HiA....11..646K} {11, 646}

\bibitem[\protect\citeauthoryear{{Ledoux}}{{Ledoux}}{1951}]{Ledoux1951}
{Ledoux} P.,  1951, \mn@doi [\apj] {10.1086/145477}, \href
  {http://cdsads.u-strasbg.fr/abs/1951ApJ...114..373L} {114, 373}

\bibitem[\protect\citeauthoryear{{Lee} \& {Baraffe}}{{Lee} \&
  {Baraffe}}{1995}]{Lee1995}
{Lee} U.,  {Baraffe} I.,  1995, \aap, \href
  {http://cdsads.u-strasbg.fr/abs/1995A%26A...301..419L} {301, 419}

\bibitem[\protect\citeauthoryear{{Lee} \& {Saio}}{{Lee} \&
  {Saio}}{1987}]{Lee1987a}
{Lee} U.,  {Saio} H.,  1987, \mnras, \href
  {http://cdsads.u-strasbg.fr/abs/1987MNRAS.224..513L} {224, 513}

\bibitem[\protect\citeauthoryear{{Lenz} \& {Breger}}{{Lenz} \&
  {Breger}}{2005}]{Lenz2005}
{Lenz} P.,  {Breger} M.,  2005, \mn@doi [Communications in Asteroseismology]
  {10.1553/cia146s53}, \href
  {http://adsabs.harvard.edu/abs/2005CoAst.146...53L} {146, 53}

\bibitem[\protect\citeauthoryear{{Marques} et~al.,}{{Marques}
  et~al.}{2013}]{Marques2013}
{Marques} J.~P.,  et~al., 2013, \mn@doi [\aap] {10.1051/0004-6361/201220211},
  \href {http://cdsads.u-strasbg.fr/abs/2013A%26A...549A..74M} {549, A74}

\bibitem[\protect\citeauthoryear{{Miglio}, {Montalb{\'a}n}, {Noels}  \&
  {Eggenberger}}{{Miglio} et~al.}{2008}]{Miglio2008}
{Miglio} A.,  {Montalb{\'a}n} J.,  {Noels} A.,   {Eggenberger} P.,  2008,
  \mn@doi [\mnras] {10.1111/j.1365-2966.2008.13112.x}, \href
  {http://cdsads.u-strasbg.fr/abs/2008MNRAS.386.1487M} {386, 1487}

\bibitem[\protect\citeauthoryear{{Mosser} et~al.,}{{Mosser}
  et~al.}{2012}]{Mosser2012b}
{Mosser} B.,  et~al., 2012, \mn@doi [\aap] {10.1051/0004-6361/201220106}, \href
  {http://adsabs.harvard.edu/abs/2012A%26A...548A..10M} {548, A10}

\bibitem[\protect\citeauthoryear{{Murphy}, {Fossati}, {Bedding}, {Saio},
  {Kurtz}, {Grassitelli}  \& {Wang}}{{Murphy} et~al.}{2016}]{Murphy2016}
{Murphy} S.~J.,  {Fossati} L.,  {Bedding} T.~R.,  {Saio} H.,  {Kurtz} D.~W.,
  {Grassitelli} L.,   {Wang} E.~S.,  2016, \mn@doi [\mnras]
  {10.1093/mnras/stw705}, \href
  {http://adsabs.harvard.edu/abs/2016MNRAS.459.1201M} {459, 1201}

\bibitem[\protect\citeauthoryear{{Ouazzani}, {Dupret}  \& {Reese}}{{Ouazzani}
  et~al.}{2012}]{Ouazzani2012b}
{Ouazzani} R.-M.,  {Dupret} M.-A.,   {Reese} D.~R.,  2012, \mn@doi [\aap]
  {10.1051/0004-6361/201219548}, \href
  {http://adsabs.harvard.edu/abs/2012A%26A...547A..75O} {547, A75}

\bibitem[\protect\citeauthoryear{{Ouazzani}, {Roxburgh}  \&
  {Dupret}}{{Ouazzani} et~al.}{2015}]{Ouazzani2015}
{Ouazzani} R.-M.,  {Roxburgh} I.~W.,   {Dupret} M.-A.,  2015, \mn@doi [\aap]
  {10.1051/0004-6361/201525734}, \href
  {http://cdsads.u-strasbg.fr/abs/2015A%26A...579A.116O} {579, A116}

\bibitem[\protect\citeauthoryear{Press, Teukolsky, Vetterling  \&
  Flannery}{Press et~al.}{1996}]{Press1996}
Press W.~H.,  Teukolsky S.~a.,  Vetterling W.~T.,   Flannery B.~P.,  1996,
  {Numerical Recipes in Fortran 77: the Art of Scientific Computing. Second
  Edition}.
 Vol. 1

\bibitem[\protect\citeauthoryear{{Reese}, {Ligni{\`e}res}  \&
  {Rieutord}}{{Reese} et~al.}{2006}]{Reese2006}
{Reese} D.,  {Ligni{\`e}res} F.,   {Rieutord} M.,  2006, \mn@doi [\aap]
  {10.1051/0004-6361:20065269}, \href
  {http://cdsads.u-strasbg.fr/abs/2006A%26A...455..621R} {455, 621}

\bibitem[\protect\citeauthoryear{{Reese}, {MacGregor}, {Jackson}, {Skumanich}
  \& {Metcalfe}}{{Reese} et~al.}{2009}]{Reese2009a}
{Reese} D.~R.,  {MacGregor} K.~B.,  {Jackson} S.,  {Skumanich} A.,   {Metcalfe}
  T.~S.,  2009, \mn@doi [\aap] {10.1051/0004-6361/200811510}, \href
  {http://cdsads.u-strasbg.fr/abs/2009A%26A...506..189R} {506, 189}

\bibitem[\protect\citeauthoryear{{Rogers} \& {Nayfonov}}{{Rogers} \&
  {Nayfonov}}{2002}]{Rogers2002}
{Rogers} F.~J.,  {Nayfonov} A.,  2002, \mn@doi [\apj] {10.1086/341894}, \href
  {http://cdsads.u-strasbg.fr/abs/2002ApJ...576.1064R} {576, 1064}

\bibitem[\protect\citeauthoryear{{Roxburgh}}{{Roxburgh}}{2006}]{Roxburgh2006}
{Roxburgh} I.~W.,  2006, \mn@doi [\aap] {10.1051/0004-6361:20065109}, \href
  {http://cdsads.u-strasbg.fr/abs/2006A%26A...454..883R} {454, 883}

\bibitem[\protect\citeauthoryear{{Royer}}{{Royer}}{2009}]{Royer2009}
{Royer} F.,  2009, in The Rotation of Sun and Stars. pp 207--230,
  \mn@doi{10.1007/978-3-540-87831-5_9}

\bibitem[\protect\citeauthoryear{{Saio}, {Kurtz}, {Takata}, {Shibahashi},
  {Murphy}, {Sekii}  \& {Bedding}}{{Saio} et~al.}{2015}]{Saio2015}
{Saio} H.,  {Kurtz} D.~W.,  {Takata} M.,  {Shibahashi} H.,  {Murphy} S.~J.,
  {Sekii} T.,   {Bedding} T.~R.,  2015, \mn@doi [\mnras]
  {10.1093/mnras/stu2696}, \href
  {http://adsabs.harvard.edu/abs/2015MNRAS.447.3264S} {447, 3264}

\bibitem[\protect\citeauthoryear{{Salmon}, {Montalb{\'a}n}, {Reese}, {Dupret}
  \& {Eggenberger}}{{Salmon} et~al.}{2014}]{Salmon2014}
{Salmon} S.~J.~A.~J.,  {Montalb{\'a}n} J.,  {Reese} D.~R.,  {Dupret} M.-A.,
  {Eggenberger} P.,  2014, \mn@doi [\aap] {10.1051/0004-6361/201323259}, \href
  {http://adsabs.harvard.edu/abs/2014A%26A...569A..18S} {569, A18}

\bibitem[\protect\citeauthoryear{{Schatzman}}{{Schatzman}}{1962}]{Schatzman1962}
{Schatzman} E.,  1962, Annales d'Astrophysique, \href
  {http://adsabs.harvard.edu/abs/1962AnAp...25...18S} {25, 18}

\bibitem[\protect\citeauthoryear{{Schmid} et~al.,}{{Schmid}
  et~al.}{2015}]{Schmid2015}
{Schmid} V.~S.,  et~al., 2015, \mn@doi [\aap] {10.1051/0004-6361/201526945},
  \href {http://adsabs.harvard.edu/abs/2015A%26A...584A..35S} {584, A35}

\bibitem[\protect\citeauthoryear{{Scuflaire}, {Th{\'e}ado}, {Montalb{\'a}n},
  {Miglio}, {Bourge}, {Godart}, {Thoul}  \& {Noels}}{{Scuflaire}
  et~al.}{2008a}]{Scuflaire2008b}
{Scuflaire} R.,  {Th{\'e}ado} S.,  {Montalb{\'a}n} J.,  {Miglio} A.,  {Bourge}
  P.-O.,  {Godart} M.,  {Thoul} A.,   {Noels} A.,  2008a, \mn@doi [\apss]
  {10.1007/s10509-007-9650-1}, \href
  {http://adsabs.harvard.edu/abs/2008Ap%26SS.316...83S} {316, 83}

\bibitem[\protect\citeauthoryear{{Scuflaire}, {Montalb{\'a}n}, {Th{\'e}ado},
  {Bourge}, {Miglio}, {Godart}, {Thoul}  \& {Noels}}{{Scuflaire}
  et~al.}{2008b}]{Scuflaire2008a}
{Scuflaire} R.,  {Montalb{\'a}n} J.,  {Th{\'e}ado} S.,  {Bourge} P.,  {Miglio}
  A.,  {Godart} M.,  {Thoul} A.,   {Noels} A.,  2008b, \mn@doi [\apss]
  {10.1007/s10509-007-9577-6}, \href
  {http://cdsads.u-strasbg.fr/abs/2008Ap%26SS.316..149S} {316, 149}

\bibitem[\protect\citeauthoryear{{Tassoul}}{{Tassoul}}{1980}]{Tassoul1980}
{Tassoul} M.,  1980, \mn@doi [\apjs] {10.1086/190678}, \href
  {http://cdsads.u-strasbg.fr/abs/1980ApJS...43..469T} {43, 469}

\bibitem[\protect\citeauthoryear{{Tkachenko} et~al.,}{{Tkachenko}
  et~al.}{2013}]{Tkachenko2013}
{Tkachenko} A.,  et~al., 2013, \mn@doi [\aap] {10.1051/0004-6361/201220978},
  \href {http://adsabs.harvard.edu/abs/2013A%26A...556A..52T} {556, A52}

\bibitem[\protect\citeauthoryear{{Unno}, {Osaki}, {Ando}, {Saio}  \&
  {Shibahashi}}{{Unno} et~al.}{1989}]{Unno1989}
{Unno} W.,  {Osaki} Y.,  {Ando} H.,  {Saio} H.,   {Shibahashi} H.,  1989,
  {Nonradial oscillations of stars}

\bibitem[\protect\citeauthoryear{{Uytterhoeven} et~al.,}{{Uytterhoeven}
  et~al.}{2011}]{Uytterhoeven2011}
{Uytterhoeven} K.,  et~al., 2011, \mn@doi [\aap] {10.1051/0004-6361/201117368},
  \href {http://adsabs.harvard.edu/abs/2011A%26A...534A.125U} {534, A125}

\bibitem[\protect\citeauthoryear{{Van Reeth} et~al.,}{{Van Reeth}
  et~al.}{2015}]{VanReeth2015}
{Van Reeth} T.,  et~al., 2015, \mn@doi [\apjs] {10.1088/0067-0049/218/2/27},
  \href {http://adsabs.harvard.edu/abs/2015ApJS..218...27V} {218, 27}

\bibitem[\protect\citeauthoryear{{Van Reeth}, {Tkachenko}  \& {Aerts}}{{Van
  Reeth} et~al.}{2016}]{VanReeth2016}
{Van Reeth} T.,  {Tkachenko} A.,   {Aerts} C.,  2016, \mn@doi [\aap]
  {10.1051/0004-6361/201628616}, \href
  {http://adsabs.harvard.edu/abs/2016A%26A...593A.120V} {593, A120}

\bibitem[\protect\citeauthoryear{{Zwintz}, {Fossati}, {Ryabchikova}, {Kaiser},
  {Gruberbauer}, {Barnes}, {Baglin}  \& {Chaintreuil}}{{Zwintz}
  et~al.}{2013}]{Zwintz2013}
{Zwintz} K.,  {Fossati} L.,  {Ryabchikova} T.,  {Kaiser} A.,  {Gruberbauer} M.,
   {Barnes} T.~G.,  {Baglin} A.,   {Chaintreuil} S.,  2013, \mn@doi [\aap]
  {10.1051/0004-6361/201220127}, \href
  {http://adsabs.harvard.edu/abs/2013A%26A...550A.121Z} {550, A121}

\makeatother
\end{thebibliography}




\bsp	
\label{lastpage}
\end{document}